\documentclass{aa}
\usepackage{graphicx}
\usepackage{natbib}
\bibpunct{(}{)}{;}{a}{}{,}

\begin{document}

\title{Integrated spectral analysis of 18 concentrated star clusters in the 
Small Magellanic Cloud
}

\author{A. E. Piatti\inst{1}, J. F. C. Santos Jr.\inst{2}, J. J. 
Clari\'a\inst{3}, \\
E. Bica\inst{4}, A. V. Ahumada\inst{3}, M.C. Parisi\inst{3}}

\institute{Instituto de Astronom\'{\i}a y F\'{\i}sica del Espacio, CC 67, Suc.
28, 1428, Buenos Aires, Argentina \and Departamento de F\'{\i}sica, ICEx, UFMG,
CP 702, 30123-970 Belo Horizonte, MG, Brazil \and Observatorio Astron\'omico,
Laprida 854, 5000, C\'ordoba, Argentina \and Depto. de Astronom\'{\i}a, UFRGS, 
CP 15051, 91500-970, Porto Alegre, Brazil}

\offprints{J. F. C. Santos Jr., \email{jsantos@fisica.ufmg.br}}

\date{Received / Accepted }

\abstract{We present in this study flux-calibrated integrated spectra in the 
range (3600-6800) \AA\ for 18 concentrated SMC clusters. Cluster reddening 
values were estimated by interpolation between the extinction maps of Burstein 
\& Heiles (1982, AJ, 87, 1165) and Schlegel et al. (1998, ApJ, 500, 525). The 
cluster parameters were derived from the template matching procedure by 
comparing the line strengths and continuum distribution of the cluster spectra 
with those of template cluster spectra with known parameters and from the 
equivalent width (EW) method. In this case, new calibrations were used together
with diagnostic diagrams involving the sum of EWs of selected  spectral lines. 
A very good agreement between ages derived from both methods was found. The 
final cluster ages obtained from the weighted average of values taken from the 
literature and the present measured ones range from 15 Mr (e.g. L\,51) to 7 
Gyr (K\,3). Metal abundances have been derived for only 5 clusters from the 
present sample, while metallicity values directly averaged from published 
values for other 4 clusters have been adopted. Combining the present cluster 
sample with 19 additional SMC clusters whose ages and metal abundances were put
onto a homogeneous scale, we analyse the age and metallicity distributions in 
order to explore the SMC star formation history and its spatial extent. By 
considering the distances of the clusters from the SMC centre instead of their 
projections onto the right ascension and declination axes, the present 
age-position relation suggests that the SMC inner disk could have been related
to a cluster formation episode which reached the peak $\sim$ 2.5 Gyr 
ago. Evidence for an age gradient in the inner SMC disk is also presented.
}

\maketitle
\markboth{A.E. Piatti et al.: Spectral analysis of 18 SMC clusters}{}

\section{Introduction}

The Small Magellanic Cloud (SMC) is a galaxy rich in star clusters of all ages 
and different types of field populations \citep{h88,h89,detal01}. An 
interesting feature in the chemical enrichment history of the SMC known up to 
now is that no very metal-poor old cluster has been observed in this galaxy 
\citep{d91,duetal01}. \citet{psc01} studied 5 outlying intermediate-age 
clusters in the SMC and, combined to other data in the literature, studied the 
age-metallicity relationship, showing that epochs of sudden chemical enrichment
take place in the age-metallicity plane. This favours a bursting star formation
history for the SMC as opposed to a continuous one. Recently, \citet{petal05} 
confirmed, with new observations, the occurrence of an important bursting star 
formation episode at $\sim$ 2.5 Gyr.

A star cluster spectral library at the SMC metallicity level can be useful for 
analyses of star clusters in dwarf galaxies observable by means of ground-based
large telescopes as well as the Hubble Space Telescope (HST). In addition, 
such metal-poor library appears to be also useful for the study of a fraction 
of star clusters in massive galaxies, due to cannibalism. Indeed, in the Milky 
Way galaxy at least four globular clusters have been accreted from the 
Sagittarius dwarf galaxy \citep{da95}, and the open clusters AM-2 and 
Tombaugh\,5 appear to be related to the Canis Major dwarf galaxy 
\citep{betal04}.

In this sense, spectral libraries of stars \citep[e.g.,][]{sc92}, open clusters
\citep{pbc02} or star clusters in general \citep{ba86} are important datasets 
for spectral classifications and extraction of parameter information for target
stars or star clusters \citep[e.g.,][]{petal02b} and galaxies 
\citep[e.g.,][]{b88}.

Samples of integrated spectra of SMC clusters were initially small, 
corresponding to the most prominent clusters such as NGC\,121, NGC\,419, 
NGC\,330 and others \citep{ba86,setal95}. \citet{aetal02} analysed integrated 
spectra in the range 3600-6800\AA\, for 16 star clusters in the SMC, estimating
ages and reddening values. That study has constituted a fundamental step 
forward towards a cluster spectral library at low metallicities.

A comprehensive catalogue of SMC clusters was produced by \citet{bs95}, and 
updated in \citet{bd00}. In it, cross-identifications for different 
designations, coordinates, angular sizes and references to previous catalogues 
are provided. The angular distribution of SMC clusters has been discussed in 
\citet{bs95}: most clusters are projected on the SMC main body and a 
significant fraction are outlyers. The line-of-sight (LOS) depth of populous 
clusters in the SMC was analysed by \citet{cetal01}, who found significant 
depth effects, with a triaxial shape of 1:2:4 for the declination, the right 
ascension, and the LOS depth of the SMC, respectively.

The present cluster sample complements previous ones, in an effort to provide 
a spectral library with several clusters per age bin. At the same time, we 
study the clusters themselves individually, determining their parameters and 
analysing the age distribution, in order to explore the SMC star formation 
history and its spatial extent. To estimate the clusters' ages, we employ the 
new calibrations and diagnostic diagrams recently provided by 
\citet[][ hereafter SP]{sp04} for visible integrated spectra, along with 
template spectra \citep[e.g.,][]{setal95,aetal02}. We confirm the reliability 
of the procedure proposed by SP in determining clusters' ages, since we 
included in the sample not only unstudied or poorly studied clusters but also 
some control clusters with well-known fundamental parameters. 

In Sec. 2 we describe the different sets of observations and the reduction 
procedure performed. The analyses of the integrated spectra through the 
template matching and equivalent width methods are developed in Sec. 3, in 
which we also include some considerations for individual clusters. In Sec. 4 we
discuss the present results in the light of the star formation history of the 
SMC. Finally, in Sec. 5 we summarize the main conclusions of this work. 

\section{Spectra acquisition and reduction}
The objects studied here are part of a systematic spectroscopic survey of SMC 
star clusters which is being undertaken at Complejo Astron\'omico El Leoncito 
(CASLEO) in San Juan (Argentina) and Cerro Tololo Inter-American Observatory 
(CTIO, Chile). The first results of this survey dealt with 16 concentrated 
star clusters \citep{aetal02}, approximately half of which constitute 
previously unstudied objects.

The observations analysed in this study were carried out with the CASLEO 2.15 m
telescope during four nights in November 2001 and five nights in October 2002 
and with the CTIO 1.5 m telescope during four nights in September 2003. In all 
the CASLEO runs we empoyed a CCD camera containing a Tektroniks chip of 1024 x 
1024 pixels attached to a REOSC spectrograph (simple mode), the size of each 
pixel being 24 $\mu$m x 24 $\mu$m; one pixel corresponds to 0.94$\arcsec$ on 
the sky. The slit was set in the East-West direction and the observations were 
performed by scanning the slit across the objects in the North-South direction 
in order to get a proper sampling of cluster stars. The long slit corresponding
to 4.7$\arcmin$ on the sky, allowed us to sample regions of the background sky.
We used a grating of 300 grooves mm$^{\rm -1}$, producing an average dispersion
in the observed region of $\approx$ 140 \AA/mm (3.46 \AA/pixel). The spectral 
coverage was $\approx$ 3600-6800 \AA. The seeing during the CASLEO nights was 
typically 2.0$\arcsec$. The slit width was 4.2$\arcsec$, providing a resolution
[full width at half-maximum (FWHM)] of $\approx$ 14 \AA, as deduced from the 
comparison lamp lines. For the CTIO observations, we used a CCD Loral 1K chip 
of 1200 x 800 pixels (pixel diameter = 15 $\mu$m), controlled by the CTIO ARCON
3.9 data acquisition system at a gain of 2.05 e$^-$ ADU$^{-1}$ with a readout 
noise of 7.4 e$^-$ ADU$^{-1}$. The same slit width of 4.2$\arcsec$ as in CASLEO
was used at CTIO, thus providing a resolution of about 11 \AA. The seeing 
during the CTIO observations was typically 1.0$\arcsec$. At least two exposures
of 30 min of each object were taken in order to correct for cosmic rays. 
Standard stars from the list of \citet{sb83} were also observed at both 
observatories for flux calibrations. Bias, darks, dome and twilight sky and 
tungsten lamp flats were taken and employed in the reductions.

The reduction of the spectra was carried out with the IRAF\footnote{IRAF is 
distributed by the National Optical Astronomy Observatories, which is operated 
by the Association of Universities for Research in Astronomy, Inc., under 
contract with the National Science Foundation} package at the Observatorio 
Astron\'omico de C\'ordoba (Argentina - CASLEO data) and at the Instituto de 
Astronom\'{\i}a y F\'{\i}sica del Espacio (Argentina - CTIO data) following the
standard procedures. Summing up, we subtracted the bias and used flat-field 
frames- previously combined - to correct the frames for high and low spatial 
frequency variations. We also checked the instrumental signature with the 
acquisition of dark frames. Then, we performed the background sky subtraction 
using pixel rows from the same frame, after having cleaned the background sky 
regions from cosmic rays. We controlled that no significant background sky 
residuals were present on the resulting spectra. The cluster spectra were 
extracted along the slit according to the cluster size and available flux. 
Five of these clusters (K\,5, K\,7, NGC\,269, K\,28 and NGC\,411) have one very
bright star located close to their main bodies. The spectra were then 
wavelength calibrated by fitting observed He-Ne-Cu (CASLEO) or He-Ar (CTIO) 
comparison lamp spectra with template spectra. The rms errors involved in these
calibrations are in average 0.40 \AA\, for both observatories. Finally, we 
applied to the cluster spectra extinction corrections and flux calibrations 
derived from the observed standard stars. We decided to use the sensitivity 
function derived from all the standard stars observed each night. This 
calibrated function turned out to be nearly the same as the nightly sensitivity
functions, but more robustly defined and with a smaller rms error. In addition,
cosmic rays on the cluster spectra were eliminated. Table 1 presents the 
cluster sample including the averaged signal-to-noise (S/N) ratios of the 
spectra.

\section{Analysis of the cluster spectra}

The cluster parameters were derived by means of two methods: the template 
matching method, in which the observed spectra are compared and matched to 
template spectra with well-determined properties \citep[e.g.][ and references 
therein]{pbc02}, and the equivalent width ($EW$) method, in which diagnostic 
diagrams involving the sum of $EW$s of selected spectral lines were employed 
together with their calibrations with age and metallicity (SP). In the first 
method, a high weight is assigned to the matching of the overall continuum, 
while in the second method the spectral lines are the observables that define 
cluster parameters. Both methods rely on the library of star cluster integrated
spectra with well-determined properties, accomplished in various studies 
\citep[e.g.][ and references therein]{ba86,pbc02} and made available through 
the CDS/Vizier catalogue database 
at http://vizier.u-strasbg.fr/cgi-bin/VizieR?-source=III/219 \citep{sab02}.

\subsection{Template matching method}

All 18 clusters in our sample are well represented by a wide variety of stellar
populations, as may be noticed from their spectra overall appearance. The 
template spectra useful for the present sample are: Yb (5-10 Myr), Yd (40 Myr),
Ye (45-75 Myr), Yg (200-350 Myr), Yh (0.5\,Gyr), Ia (1\,Gyr) and Ib (3-4\,Gyr),
which represent young and intermediate-age populations built from Galactic open
clusters  \citep{pbc02}, and G3 ($>10$\,Gyr, [Fe/H] = -1.0), G4 ($>10$\,Gyr, 
[Fe/H] = -1.5) and G5 ($>10$\,Gyr, [Fe/H] = -2.0), which represent old stellar 
populations built from Galactic globular clusters \citep{b88}.

The template matching method consists of achieving the best possible match 
between the analysed cluster spectrum and a template spectrum of known age and 
metallicity. In this process we selected, among the available template spectra,
the ones which minimize the flux residuals, calculated as the difference 
(cluster - template)/cluster. Note that differences between the cluster and 
template spectra are expected to be found due to variations in the stellar 
composition of the cluster, such as the presence of a relatively bright 
star with particular spectral features or contamination of a field star close
to the direction towards the cluster. 

Since the continuum distribution is also affected by reddening, we first 
adopted a colour excess $E(B-V)$ for each cluster, taking into account the 
\citet[][ hereafter BH]{bh82} and \citet[][ hereafter SFD]{sfd98} extinction 
maps, and then corrected the observed spectra accordingly before applying the 
template match method. We recall that one can deredden an integrated spectrum 
and simultaneously estimate the cluster age. However, in order to make the age 
estimate more robust, we preferred to match reddening corrected cluster spectra
with template spectra. Thus, instead of having to handle two variables in the 
match (reddening and age), we limit it to find only the cluster age. 

The maps of BH and SFD are frequently used to estimate the colour excesses of 
clusters located in the direction towards the Magellanic Clouds \citep[see, 
e.g.,][]{psc01,duetal01}. SFD found that at high-latitude regions, their dust
maps correlate well with maps of H\,I emission, but deviations are coherent in 
the sky and are especially conspicuous in regions of saturation of HI emission 
towards denser clouds and of formation of H$_2$ in molecular clouds. The SMC 
is quite transparent, the average foreground and internal reddenings being 0.01
and 0.04, respectively \citep{duetal01}. The typical reddening towards the SMC 
estimated from the median dust emission in annuli surrounding the galaxy is 
$E$($B-V$)$=0.037$ (SFD). Therefore, we assume that relatively high SFD values 
are saturated and we then use the BH values. For clusters with non saturated 
SFD values, the difference between SFD and BH colour excesses resulted in, at 
the most, 0.02 mag; the SFD zero-point being made consistent with the BH 
maps by subtracting 0.02 mag in $E(B-V)_{\rm SFD}$. The results are shown in 
Figs. 1 to 18.

\subsection{Equivalent width method}

Before measuring $EW$s in the observed spectra, they  were set to the 
rest-frame according to the Doppler shift of H Balmer lines. Next, the spectra 
were normalized to F$_{\lambda}=1$ at 5870 \AA~ and smoothed to the typical 
resolution of the database ($\approx$ 10-15 \AA).
 
Spectral fluxes at 3860, 4020, 4150, 4570, 4834, 4914 and 6630 \AA~ were used 
as guidelines in order to define the continuum according to \cite{ba86}. The 
$EW$s of H Balmer,  K\,Ca\,II, G band (CH) and Mg\,I (5167 + 5173 + 5184 \AA) 
were measured within the spectral windows defined by \cite{ba86} and using IRAF
task {\it splot}. Boundaries for the K\,Ca\,II, G band (CH), Mg\,I, H$\delta$, 
H$\gamma$ and  H$\beta$ spectral windows are, respectively, (3908-3952) \AA, 
(4284-4318) \AA, (5156-5196) \AA, (4082-4124) \AA, (4318-4364) \AA, and 
(4846-4884) \AA. Such a procedure has been applied consistently making the 
$EW$s from integrated spectra safely comparable with the well-known cluster 
database. Table 2 presents these measurements as well as the sum of $EW$s of 
the three metallic lines ($S_m$) and of the three Balmer lines H$_{\delta}$, 
H$_{\gamma}$ and H$_{\beta}$ ($S_h$). $S_m$ and $S_h$ are shown to be useful in
the discrimination of old, intermediate-age and young systems 
\citep[][ SP]{r82,dbc99}. Typical errors of $\approx$ 10 \% on individual $EW$ 
measurements were obtained by tracing slightly different continua. By using 
the sums of $EW$s $S_h$ and $S_m$  separately, the $EW$ relative errors are 
lowered ($\approx7\%$ smaller range than the individual $EW$ errors), improving
their  sensitivity to cluster age and metallicity (SP).

The sums of $EW$s $S_h$ and $S_m$ presented in Table 2 were used to estimate 
cluster parameters according to their calibrations as a function of age and 
metallicity given by SP. Such calibrations are based on visible integrated 
spectra of Galactic and Magellanic Cloud clusters for which age and metallicity
were well-determined and put within homogeneous scales. In summary, the 
calibrations, aided by diagnostic diagrams involving $S_m$ and $S_h$, allow 
one to obtain age for star clusters younger than $\approx$ 10 Gyr and 
metallicity for older ones. Yet, a degeneracy occurs for globular age-like 
clusters with $[Fe/H] > -1.4$ and intermediate-age clusters 
($2.5$ $<$ $t$ (Gyr) $<$ $10$), which cannot be discriminated using this 
method. In this case, it is necessary to constrain age by using an independent 
method (e.g., the template matching one) and then obtain metallicity with the 
SP's calibration, if the cluster is old.  It is worth mentioning that
only 5 SMC clusters are included in the SP's calibration, but since they
follow the general trend of Galactic clusters in the diagnostic diagrams,
we judged safe to apply that calibration to the present sample.
The derived ages and metallicities for the cluster sample are summarized in 
Table 3. In columns 6 and 9, the methods used to obtain age and metallicity 
are indicated. Except for K\,28, with a low S/N spectrum, all remaining 
clusters were age-ranked according to the $EW$ method based on $S_h$ and $S_m$ 
measurements. In the case of NGC\,269, we decided to use only $S_m$, since the 
substraction of the spectrum of the symbiotic nova SMC\,3 could affect the 
$EW$s of the cluster H Balmer lines (see details in Section 3.3.7). The 
template method was applied to the whole sample either independently from the 
$EW$ method (minus sign in column 6) or in conjunction with the $EW$ method 
(plus sign in column 6). Note that we only had to employ template and $EW$s 
methods in conjunction for clusters in the age-metallicity degeneracy range. 
We found a very good agreement between ages derived from both methods. The 
final cluster ages obtained from the weighted average of values taken from the 
literature (columns 3 and 4) and the measured present ones ($t_{\rm m}$)
are listed in column 7. Their respective errors take into account the 
dispersion of the values averaged and/or the estimated uncertainties for 
$t_{\rm m}$. Column 2 lists the colour excesses adopted for the 
clusters. 

The last two columns of Table 3 show the cluster metallicities adopted whenever
possible and their corresponding sources, respectively. Some clusters have 
metal abundances directly averaged from published values. For K\,3, we used 
eq. (8) of SP. Three clusters (L\,5, K\,5 and K\,28) have metallicities 
derived from a technique involving morphological features in the cluster 
colour-magnitude diagram (CMD) \citep{petal02b,petal05}, which we corrected 
for age degeneracy using the present ages. Finally, we fitted Padova isochrones
\citep{gbb02} to the K\,6 CMD obtained by \cite{mrb02} and yielded a cluster 
metallicity of [Fe/H] = -0.7, assuming for the cluster the reddening and age of
Table\,3 and the SMC apparent distance modulus $(m-M)$ = 19.0 \citep{cetal00}. 
The fit was performed on an extracted CMD containing stars distributed around 
2$\arcmin$ from the cluster centre, with the aim of avoiding field star 
contamination.

\subsection{Individual cluster analysis}

We have revised the literature on the cluster parameters below. More weight has
been assigned to ages determined from isochrone fitting to CMD data, but when 
such information was not available, ages based on integrated indices were also 
considered. No previous age information was found either for HW\,8 (Fig. 6) or
IC\,1641 (Fig. 17).

\subsubsection{L\,5}

\cite{petal05} have derived $t=4.3$\,Gyr and $[Fe/H]=-1.2$ for this cluster. 
Much like K\,5, the age of L\,5 has been estimated to be 0.8\,Gyr according to 
both methods employed in the present work. A correction to the metallicity 
provided by \cite{petal05} revised it to $[Fe/H]=-1.1$ for its significantly 
younger age. Fig. 1 shows the best template combination for L\,5, i. e., the 
average of Ia and Yh templates with a reddening of $E(B-V)$ = 0.03. This is the
cluster with the most discrepant age in the sample with respect to the 
published cluster ages. We did not find any reason for such difference, apart 
from a relative low S/N ratio in the observed spectrum.

\subsubsection{K\,5}

\cite{bdp86} derived for K\,5 the following parameters from integrated 
photometry of the H$\beta$ and G band absorption features: $[Z/Z_\odot] = -1.1 
\pm 0.2$ and $t = 3.2 \pm 0.3$\,Gyr, while the recent study by \cite{petal05} 
yields $[Fe/H] = -0.6$ and $t = 2.0$\,Gyr. The template method estimate 
for K\,5 age is $t = 0.8$\,Gyr, according to its spectral resemblance to an 
average of templates Ia and Yh, after applying a reddening correction of E(B-V)
= 0.02 (Fig. 2). Its metallicity has been corrected to $[Fe/H]=-0.5$, 
following an age revision on the \cite{petal05} value.

\subsubsection{K\,3}

\cite{rcm84} determined an age of 5-8 Gyr from BR photometry and isochrone 
fitting. K\,3 was included in the integrated photometric study by \cite{bdp86},
who derived $t\geq$ 10\,Gyr and $[Z/Z_{\odot}] = -1.5 \pm 0.2$. \cite{msf98}
obtained $[Fe/H] = -1.16 \pm 0.09$, $t = 6.0 \pm 1.3$\,Gyr and $E(B-V)$ = 0.0 
from HST observations and morphological parameters defined in the CMD. More 
recently, \cite{bdm01} presented a HST CMD of K\,3 making available its 
photometry, on which we have superimposed Padova isochrones \citep{gbb02} to 
obtain essentially the same parameters as those derived by \cite{msf98}. 
In the present study, an intermediate age for K\,3 is confirmed, being 
this the oldest cluster in the present sample. The template matching method 
gives for this cluster $\approx7$\,Gyr as a result of averaging the G3 and Ia 
templates (Fig. 3). Both age and metallicity obtained in the present analysis 
show good agreement with results from previous studies.

\subsubsection{K\,6}

From CCD $BV$ photometry selected for an inner region (r $<$ 35$\arcsec$) of 
K\,6, \cite{mrb02} derived an age of 1-1.3 Gyr for this cluster. The spectrum 
comparison leads to a match of K\,6 spectrum with the template Ib (3-4\,Gyr), 
combined with a reddening correction of $E(B-V)$ = 0.03 (Fig. 4). However, a 
smaller age is suggested by the $EW$ method, being $t$ = 1.6 Gyr the final 
adopted value. By fitting Padova isochrones \citep{gbb02} to the CMD data of 
\cite{mrb02} and assuming the above mentioned age and the apparent distance 
modulus $(m-M)$ = 19 \citep{cetal00}, an estimate of the cluster metallicity 
was also obtained, i. e., [Fe/H] = -0.7.

\subsubsection{K\,7}

\cite{mjd82} carried out CCD $BR$ photometry of K\,7 obtaining $t$ = 3.5 $\pm$ 
1 Gyr by isochrone fitting with $E(B-V)$ = 0.04. The template spectrum Ib 
(3-4\,Gyr) was initially tried as a match to the K\,7 spectrum, but its redder 
colour cannot be accounted for by a large reddening correction exclusively. 
\cite{mjd82} pointed out the presence of two carbon stars close to the cluster 
centre, which are the probable contributors to the red appearance of its 
integrated spectrum. In order to check whether this is the case, a combination 
of the Ib spectrum with a carbon star spectrum taken from \cite{bsk96} spectral
library was tried. Specifically, the spectrum of the nearly solar metallicity
carbon star BM\,Gem \citep{ai00} was used in the analysis. According to our 
observations, in the cluster spatial profile the presence of the bright star 
stands out over the bulk of the cluster light. We then extracted the integrated
spectrum of the cluster plus the carbon star and of the carbon star spectrum 
alone. The flux ratio at 5870 \AA~ between the carbon star spectrum and 
the integrated one turned out to be 0.35. As a matter of fact, there is a good 
match to K\,7 spectrum if the template Ib is combined with the carbon star in 
a proportion of 65\% and 35\% of the total light at 5870\AA, respectively, and 
the resulting spectrum is reddening corrected by $E(B-V)$ = 0.02 (Fig. 5). 
Relatively large residual spectral differences still remain between the 
spectra, which may be attributed to the higher metallicity of the carbon star 
employed as template. No metallicity has been estimated for this cluster.

\subsubsection{NGC\,269}

This is an interesting case in which there is a bright emission line star 
contributing significantly to the cluster integrated spectrum. Such a 
situation, which we had found in previous cluster observations 
\citep[e.g.][]{setal95}, has been successfully treated by subtracting the star 
spectrum from the total integrated one, leaving a spectrum which better 
represents the cluster average population. Although such a procedure introduced
noise in the resulting spectrum, it allowed us to estimate the cluster age 
using the template matching method. The bright star in NGC\,269 spectrum is 
SMC\,3, a symbiotic nova composed by a M0 giant and a white dwarf orbiting 
each other in a period of $\approx4$\,years \citep{k04}. Its spectrum 
was published in the spectrophotometric atlas of \cite{mz02}. The OGLE database
includes a CMD for this cluster (OGLE-CL-SMC0046), although an age estimate 
was not provided there \citep{puk98,pu99}. This CMD shows that SMC\,3 is 
$\approx$ 2 mags brighter in $V$ than the next bright star in the cluster. In 
detail, the procedure adopted was to subtract the total integrated spectrum 
from a scaled SMC\,3 spectrum by assuming that all the emission present in the 
integrated spectrum is due to SMC\,3. In this manner, the difference between 
the spectra which minimizes the emission line residuals was obtained 
when the star contributes with 60\% of the total flux at 5870 \AA. Since the 
spectra were observed at different epochs and SMC\,3 is variable, the small but
clearly visible residuals reflect such irregularities. Another point that 
allows one to check the reliability of this procedure is the fact that 
absorption molecular bands present in SMC\,3 spectrum almost disappear in the 
resulting spectrum. The subtracted spectrum was then submitted to the template 
matching method (Fig. 7), being similar to an average of the templates Yh and 
Ia (750 Myr). Such an age is in agreement with the clusters SWB type III-IV 
\citep{swb80}. \cite{glb04} have assigned an age of 500 Myr to NGC 269, based 
on the integrated colour parameterization (``s'' parameter) by \cite{ef88}. 
However, it should be kept in mind that \cite{glb04} age ranking is intended 
to group clusters of similar integrated properties and their age groups 
encompass wide age ranges.

\subsubsection{L\,39}

The OGLE database includes a CMD for this cluster (OGLE-CL-SMC0054), with an 
isochrone based age estimate of 100 $\pm$ 23 Myr \citep{puk98,pu99}. A new 
age estimate based both on a different areal extraction of the same data and 
on the same isochrones revises it down to 80 $\pm$ 20 Myr \citep{odbd00}. 
According to \cite{glb04}, L\,39 is similar to 50 Myr old clusters. We have 
found that the cluster is 15 $\pm$ 10 Myr old, its steep continuum resembling 
those of L\,51 and L\,66 (Fig. 8).

\subsubsection{K\,28}

\cite{psc01} obtained CCD Washington photometry for this cluster deriving 
[Fe/H] = -1.45 $\pm$ 0.13 and $t$ = 2.1 $\pm$ 0.5 Gyr, with a reddening within
the range 0.06 $<$ $E(B-V)$ $<$ 0.16. In the present analysis, we have not 
applied the $EW$ method to derive parameters for K\,28 because its integrated 
spectrum has low S/N ratio, although it still seems to be adequate to the 
template matching method. Indeed, using the latter, we have 
got a good match for the template Ia (1\,Gyr) combined with a reddening of 
$E(B-V)$ = 0.06 (Fig. 9). By revising down the age obtained by \cite{psc01}, a 
corrected metallicity of [Fe/H] = -1.0 was derived.

\subsubsection{NGC\,294}
This cluster (OGLE-CL-SMC0090) has a CMD included in the OGLE database. Its 
estimated age based on the isochrone fitting method is 316 $\pm$ 73 Myr 
\citep{puk98,pu99}. The study published by \cite{odbd00} gives 300 $\pm$ 50 
Myr. Since there is no SWB type assigned to this cluster, \cite{glb04} 
included NGC\,294 in their 1\,Gyr cluster group due to the similarity of its 
integrated colours to the colours of  SWB\,IV clusters. This age seems to be 
in disagreement with the previously mentioned works and also with ours, which 
gives $t$ = 300 $\pm$ 100 Myr (Fig. 10). Visible cluster images do not show 
any bright star in the cluster core, and therefore the age discrepancy cannot 
be due to sampling effects.

\subsubsection{L\,51}
This cluster has spectral similarities with L\,66, for which a CMD is available
(see below). \cite{glb04} have assigned an age of 10 Myr to L\,51 based on the
``s'' parameter \citep{ef88}, in agreement with our estimate of $t$ = 15$\pm$ 
10 Myr (Fig. 11).

\subsubsection{K\,42}
The OGLE database includes a CMD for this cluster (OGLE-CL-SMC0124), with an 
isochrone based age estimate of 39.8$ \pm$ 9.2 Myr \citep{puk98,pu99}. The 
study by \cite{odbd00} obtained for it a younger age, 20 $\pm$ 10 Myr, which 
seems too low compared to our estimate, $t$ = 45 $\pm$ 15 Myr (Fig. 12).

\subsubsection{L\,66}
The OGLE database includes a CMD for this cluster (OGLE-CL-SMC0129), with an 
isochrone based age of 20.0 $\pm$ 4.6 Myr \citep{puk98,pu99}. This result 
is comparable to our age estimate of $t$ = 15 $\pm$ 10 Myr (Fig. 13).

\subsubsection{NGC\,411}

The spectral features and continuum slope of NGC\,411 are comparable to the 
template spectrum Ia (1\,Gyr), when a reddening correction of $E(B-V)$ = 0.03 
is applied to the observed spectrum (Fig. 14). Although this age estimate is 
lower than that obtained by \cite{bdp86}, i.e., 3.4 $\pm$ 0.3 Gyr (and 
$[Z/Z_{\odot}] = -1.3 \pm 0.2$), the present result agrees with the study by
\cite{lr03} involving integrated spectroscopy at a higher resolution who 
obtained $t$ = 1.2 $\pm$ $0.2$ Gyr and [Fe/H] = -0.43 $\pm$ 0.14. In addition, 
two studies involving isochrone fittings to CMDs yield similar results: 
\cite{dm86} obtain $t$ = 1.5 $\pm$ 0.5 Gyr and $[Fe/H]$ = -0.9 $\pm$ 0.3, 
adopting a reddening of $E(B-V)$ = 0.04 and \cite{as99} determined  $t$ = 1.4 
$\pm$ 0.2 Gyr and $[Fe/H]$ = -0.68 $\pm$ 0.07, based on HST data. By using CCD
Washington photometry, a metallicity of $[Fe/H]$ = -0.84, which agrees with 
the previous, more recent estimates, was derived for this cluster by 
\cite{petal02b}.

\subsubsection{NGC\,419}
An age lower limit of 1 Gyr was obtained for this cluster (OGLE-CL-SMC0159) by
isochrone fitting to OGLE data \citep{puk98,pu99}. This age limit is in 
accord with the other recent age estimates by \cite{retal00}, who obtained 2.0 
$\pm$ 0.2 Gyr by means of isochrone fitting to the cluster HST CMD, and by 
\cite{dhm84}, who estimated 1.2 $\pm$ 0.5 Gyr by using isochrone fitting to 
photographic CMDs. The template matching method yields $t$ = 0.75 Gyr (Fig. 
15), but when this value is combined with an independent estimate from the 
$EW$ method, the age converges to 1.2 $\pm$ 0.4 Gyr, in close agreement with 
the result found by \cite{dhm84}. This age is consistent with [Fe/H] = -0.7 
\citep{petal02b}.

\subsubsection{NGC\,422}
As far as we know, the only age information on this cluster is based on its 
integrated colours. \cite{glb04} have considered NGC\,422 as a member of the 
50 Myr group (SWB type II), which is younger than our estimate of $t$ = 400 Myr
(Fig. 16), based on the template method. The final adopted age for NGC\,422 is 
$t$ = 300 $\pm$ 100 Myr, which results from the independent methods employed 
in the present work.

\subsubsection{NGC\,458}
Age determinations of NGC\,458 based on isochrone fitting to the cluster CMD 
were made by \cite{ps88}, \cite{sch92} and more recently by \cite{aetal03}, 
who obtained 300 Myr, 100 Myr and 140 Myr, respectively. Independent age 
estimates using the template matching (Fig. 18) and the $EW$ methods give 
$t$ = 50 Myr. Taking into account the previous literature determinations, a 
final value of $t$ = 130 $\pm$ 60 Myr was adopted, which is consistent with 
[Fe/H] = -0.23 \citep{petal02b}.

\section{Discussion}

Fig. 19 shows the positions of the studied clusters (crossed boxes) relative to
the SMC optical centre (cross), assumed to be placed (J2000) at: 00$^h$ 52$^m$ 
45$^s$, -72$\degr$ 49$\arcmin$ 43$\arcsec$ \citep{cetal01}. For the sake of 
completeness, we included 19 additional clusters (triangles) taken from Table 4
of \cite{petal02b} and studied by \cite{petal05}, which have ages and 
metallicities put onto a homogeneous scale. The collection of these 37 objects 
constitutes at the present time the largest sample of SMC clusters used to 
address the issue of the galaxy chemical  evolution. 
Thus, the results derived from this sample are valuable in the sense
that they give us the opportunity to have some clues about the galaxy history,
which obviously needs later confirmation from a larger database.
Besides the SMC Bar, 
represented by a straight line in Fig. 19, we traced two ellipses centred at 
the SMC optical centre  with their major axes aligned with the galaxy Bar. We 
adopted a $b/a$ ratio which equals to 1/2. The semi-major axes of the ellipses 
drawn in the figure have 2$\degr$ and 4$\degr$, respectively. Note that this 
elliptical geometry matches the space distribution of clusters more properly 
than a circular one. 

When describing the cluster age and metallicity distributions, the 
interpretation of the results can depend on the spatial framework used. For 
example, one can adopt as a reference system the one corresponding to the right
ascension and declination axes, or that centred on the galaxy with a coordinate
axis parallel to the Bar. Thus, if there existed an abundance gradient from the
centre and along the SMC Bar, its projection to the right ascension and 
declination axes would appear steeper. Similarly, it could be possible to 
affirm the existence of features which are actually the result of projection 
effects on these directions. By considering the distances of the clusters from 
the SMC centre instead of their projections onto the right ascension and 
declination axes, the genuine cluster age and metal abundance variations can 
be traced. Moreover, although it may be advantageous to plot ages and 
metallicities as a function of the distance from the galaxy centre, these 
plottings can result even more meaningful when the spatial variable reflects 
the flattening of the system. In the case of the SMC, this can be accomplished 
by using ellipses instead of circles around the SMC centre.

In order to examine how the cluster ages vary in terms of the distances from 
the SMC centre, we computed for each cluster the value of the semi-major axis 
($a$) that an ellipse would have if it were centred at the SMC centre, had a 
$b/a$ ratio of 1/2, and one point of its trajectory coincided with the cluster 
position. Fig. 20 shows the result obtained, in which we used the same symbols 
as in Fig. 19. The figure reveals that there are very few clusters younger than
4 Gyr in the outer disk, defined as the portion of the SMC disk with $a$ $\ge$ 
3.5$\degr$. Conversely, it would appear that there are very few clusters older 
than 4 Gyr in the inner disk. Furthermore, in the inner disk, the older the 
clusters, the larger their corresponding semi-major axes, which astonishingly 
suggests the possibility that the clusters were formed outside in, like in a 
relatively rapid collapse. As far as we are aware, this is the first time such 
an evidence is presented.

\cite{hz04} recently determined the global star 
formation and chemical enrichment history of the SMC within the inner 
4$\degr$x4.5$\degr$ area of the main body, based on UBVI photometry 
of $\sim$ 6 million 
stars from their Magellanic Clouds Photometric Survey \citep{zht97}. 
Among other results, they found that there was a rise in the mean 
star formation rate during the most recent 3 Gyr punctuated by bursts at 
2.5 Gyr, 400 Myr, and 60 Myr. The two  older events  coincide with past 
perigalactic passages of the SMC around the Milky Way 
\citep[see, e.g.,][]{ljk95}. In addition, \cite{hz04} derived a 
chemical enrichment history 
in agreement with the age-metallicity relation of the SMC clusters 
and  field variable stars. This chemical enrichment history is consistent 
with the model of \cite{pt99}, lending further support to 
the presence of a long quiescent period (3 $<$ age(Gyr) $<$ 8.4) in the SMC 
early history. Piatti et al. (2005) confirmed that $\sim$ 2.5 Gyr ago the SMC 
reached the peak of a burst of cluster formation, which corresponds to a 
very close encounter with the LMC according to recent dynamic models of 
\cite{bekki04}. It would seem reasonable, therefore, to accept that 
the burst which took place 2 Gyr ago formed both clusters and stars 
simultaneously. Particularly, the 2.5 Gyr star burst appears to have an 
annular structure and an inward propagation spanning $\sim$ 1 Gyr \citep{hz04}.

\cite{petal05} studied 10 clusters mainly located in the southern half of the 
SMC with ages and metallicities in the ranges 1.5 - 4 Gyr and -1.3 $<$ [Fe/H] 
$<$ -0.6, respectively. They also favoured a bursting cluster formation 
history as opposed to a continuous one for the SMC. The age-position relation 
shown in Fig. 20 for clusters younger than 4 Gyr adds, if it is confirmed, a 
new piece of evidence to the bursting conception of cluster formation. In the 
case of the cluster formation episode peaking at $\sim$ 2.5 Gyr 
\citep{petal05}, the burst could have triggered the formation process which 
continued producing clusters from the outermost regions to the innermost 
ones in the inner SMC disk. On this basis, the inner disk could have been 
formed during this period.

The distribution of the cluster metal abundances as a function of the distances
from the SMC centre is depicted in Fig. 21, where we used the same symbols as 
in Fig. 19. Note that in the outer disk, there are no clusters with 
iron-to-hydrogen ratios larger than [Fe/H] = -1.2, with only one exception. On 
the other hand, the inner disk is shared by both metal-poor and metal-rich 
clusters, the averaged metallicity being clearly larger 
than that for the outer disk. We thus confirm the existence of a metal 
abundance gradient for the SMC disk, in the sense that the farther a cluster 
from the galaxy centre, the poorer its metal content. However, all the clusters
with [Fe/H] $>$ -1.2 in the inner disk were formed during the last 4 Gyr, 
whereas the metal-poor ones are as old as those in the outer disk (see Fig. 
20). Consequently, the abundance gradient seems to reflect the combination 
between an older and more metal-poor population of clusters spread 
throughout the SMC and a younger and metal-richer one mainly formed in the 
inner disk. Note that some few clusters were also formed in the inner disk 
with [Fe/H] $\sim$ -1.2 (Fig. 21). We also recall that the present cluster 
sample follows the age-metallicity relation discussed in a previous work
\citep[][ see their Figure 6]{petal05}.

\section{Concluding remarks}
As part of a systematic spectroscopic survey of star clusters in the SMC, we 
present and analise in the current paper flux-calibrated integrated spectra of 
18 concentrated star clusters which, with a few exceptions, lie within the 
inner parts of the SMC. The sample of SMC clusters studied by means of 
integrated spectroscopy has now been considerably increased. Therefore, the 
present cluster spectral library at the SMC metallicity level can be useful 
for future analyses of star clusters in dwarf galaxies as well as for the 
study of a fraction of star clusters in massive galaxies.

E(B-V) colour excesses were derived for the present cluster sample by 
interpolation between the extinction maps published by \cite{bh82} and by 
\cite{sfd98}. Using template spectra with well determined cluster properties 
and equivalent widths (EWs) of the Balmer and several metallic lines, we 
determined ages and, in some cases, metallicities as well. For the SMC 
clusters HW\,8 and IC\,1641, the ages have been determined for the first time, 
while for the rest of the studied sample the ages derived from the template 
matching and EW methods exhibit very good agreement. Metal abundances have been
derived for five clusters (L\,5, K\,5, K\,3, K\,6 and K\,28), while we have 
adopted averaged metallicities from published values for other 4 clusters 
(K\,7, NGC\,411, NGC\,419 and NGC\,458). By combining the present cluster 
sample with 19 additional SMC clusters with ages and metallicities in a 
homogeneous scale, we analise the age and metallicity distributions in 
different regions of the SMC to probe the galaxy chemical enrichment and its 
spatial distribution. Very few clusters younger than 4 Gyr are found in the 
outer disk and, conversely, very few clusters older than 4 Gyr lie in the inner
disk. Furthermore, the present age-position relation for the SMC clusters in 
the inner disk suggests not only the possibility that the clusters were formed 
outside in, like in a relatively rapid collapse, but also that the inner disk 
itself could have been formed during a bursting formation mechanism, with an 
important cluster formation event centred at $\sim$ 2.5 Gyr. 
According to the recent 
results obtained by \cite{hz04}, this cluster burst, which 
occurred $\sim$ 2.5 Gyr ago, is clearly related to an episode of enhanced star 
formation having taken place about the same time ago.
Evidence is also 
presented on the existence of a radial metal abundance gradient for the SMC 
disk, which reflects the combination between an older and more metal-poor 
population of clusters distributed throughout the SMC and a younger and 
metal-richer one mainly formed in the inner disk.
 
\begin{acknowledgements}

We are grateful for the use of the CCD and data acquisition system at CASLEO, 
supported under US National Science Foundation (NSF) grant AST-90-15827. This 
work is based on observations made at CTIO, which is operated by AURA, Inc., 
under cooperative agreement with the NSF. We thank the staff members and 
technicians at CASLEO and CTIO for their kind hospitality and assistance 
during the observing runs. We gratefully acknowledge financial support from 
the Argentinian institutions CONICET, Agencia Nacional de Promoci\'on 
Cient\'{\i}fica y Tecnol\'ogica (ANPCyT) and Agencia C\'ordoba Ciencia. We 
thank Dr. Munari for sending us the spectrum of SMC\,3. This work was also
partially supported by the Brazilian institution FAPEMIG and CNPq.

\end{acknowledgements}

\clearpage

\begin{figure}
\resizebox{\hsize}{!}{\includegraphics{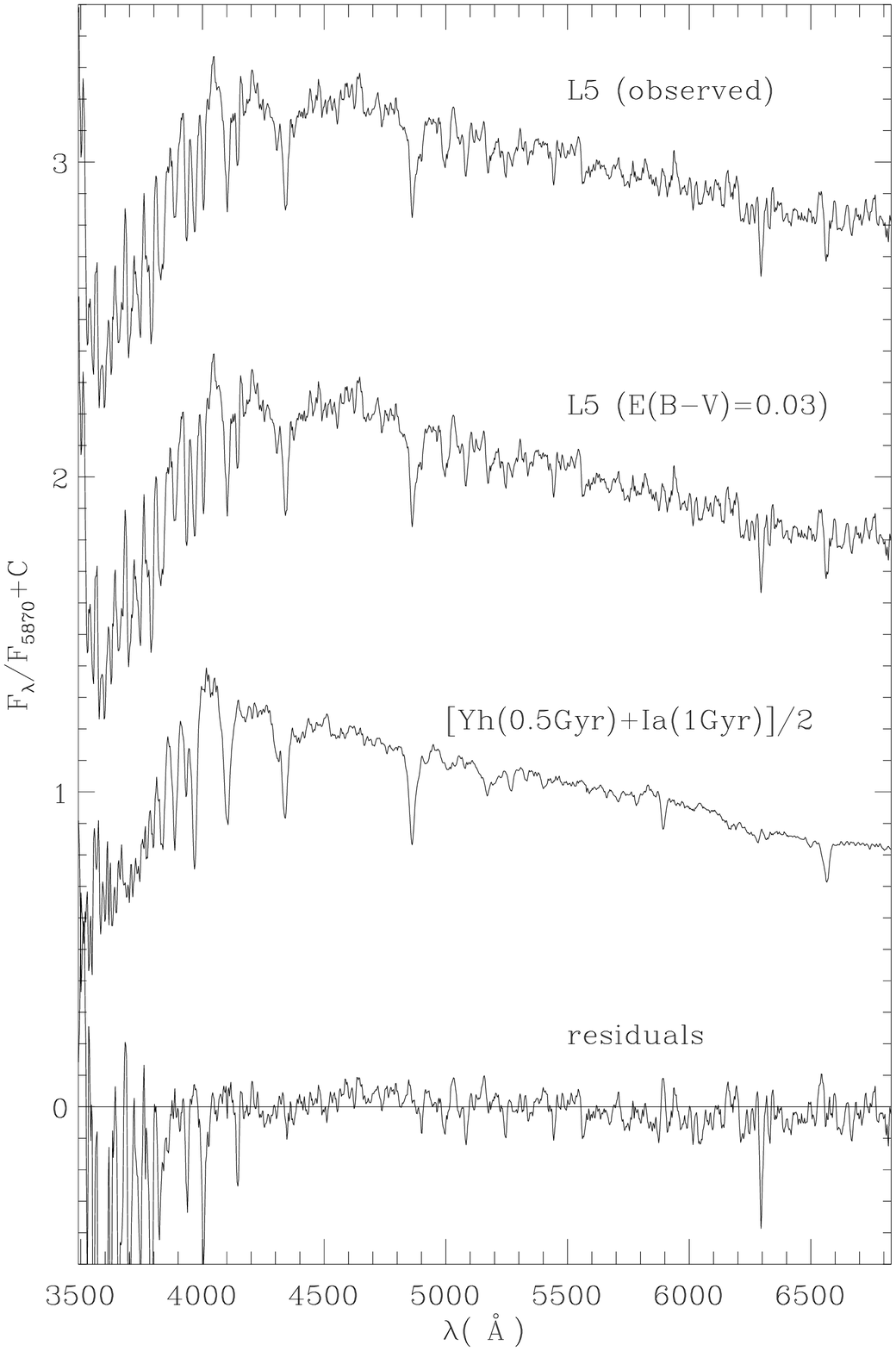}}
 \caption{Observed integrated spectrum of L\,5 (top), the spectrum corrected
for the adopted reddening $E(B-V)$ and the template spectrum which best
matches it (middle), and the residuals between both (bottom). See details 
in Sect. 3.3.1.}
\label{fig1}
\end{figure}

\begin{figure}
\resizebox{\hsize}{!}{\includegraphics{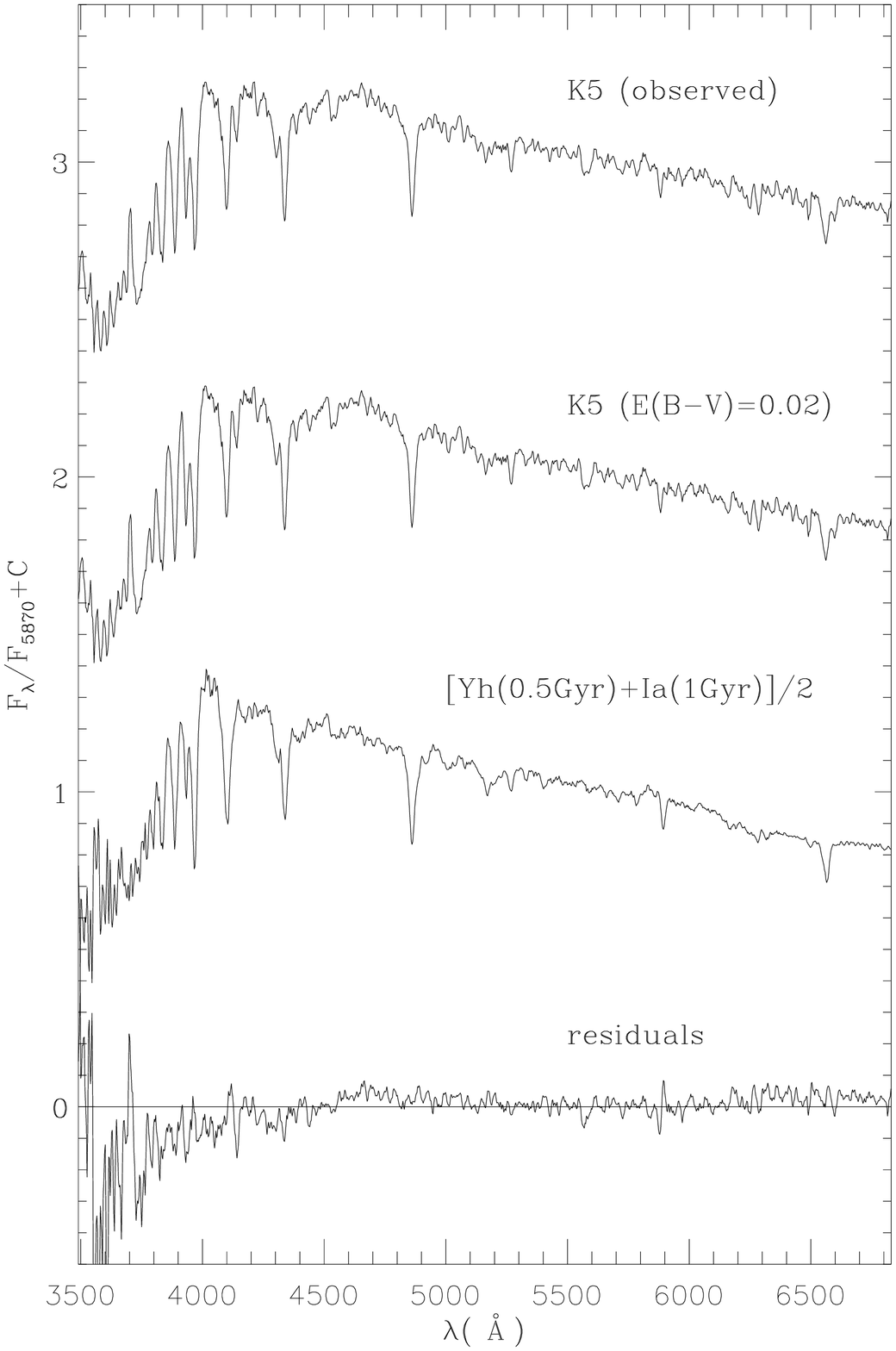}}
 \caption{Observed integrated spectrum of K\,5 (top), the spectrum corrected
for the adopted reddening $E(B-V)$ and the template spectrum which best
matches it (middle), and the residuals between both (bottom). See details 
in Sect. 3.3.2.}
\label{fig2}
\end{figure}

\begin{figure}
\resizebox{\hsize}{!}{\includegraphics{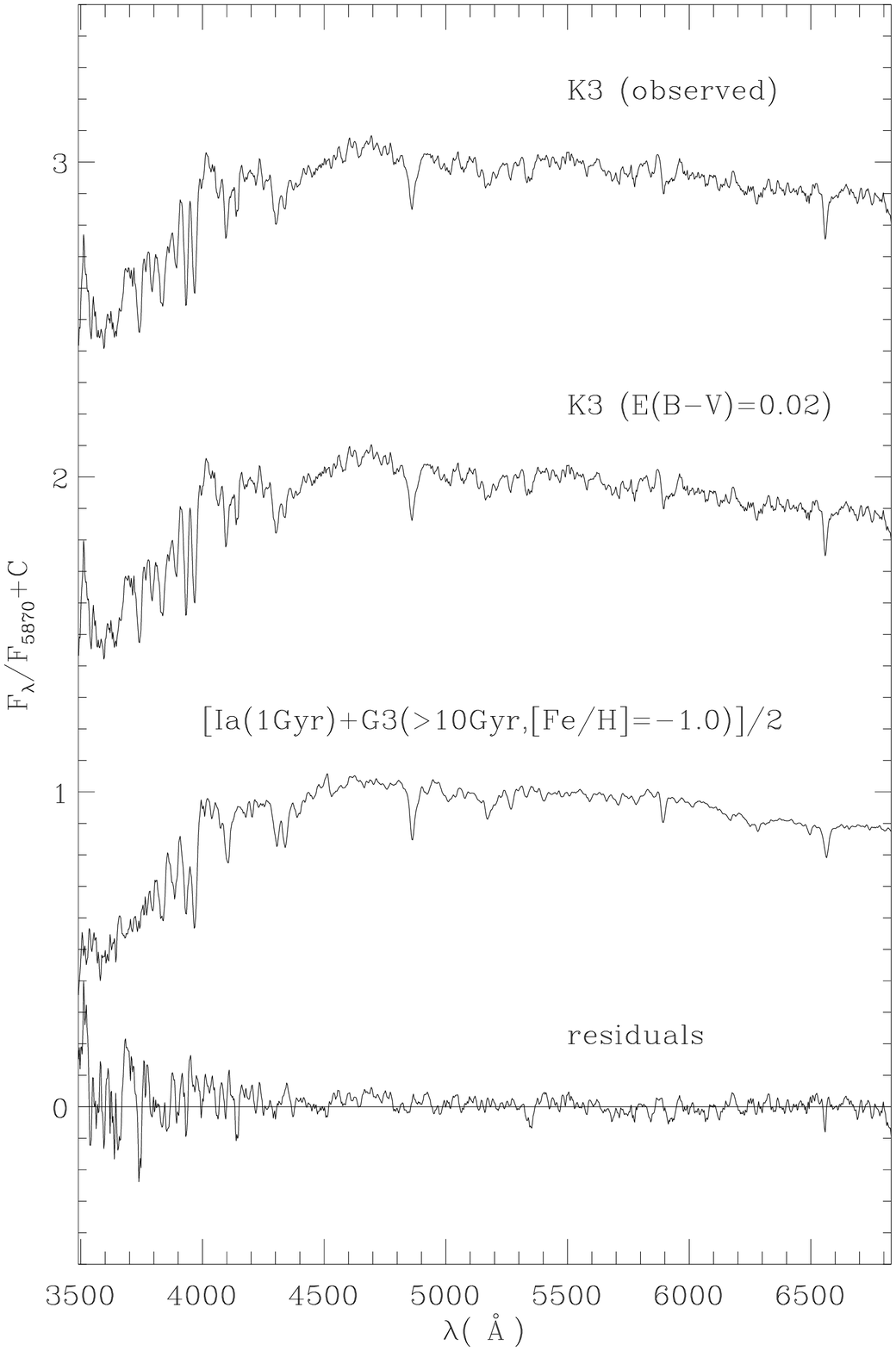}}
 \caption{Observed integrated spectrum of K\,3 (top), the spectrum corrected
for the adopted reddening $E(B-V)$ and the template spectrum which best
matches it (middle), and the residuals between both (bottom). See details 
in Sect. 3.3.3.}
\label{fig3}
\end{figure}

\begin{figure}
\resizebox{\hsize}{!}{\includegraphics{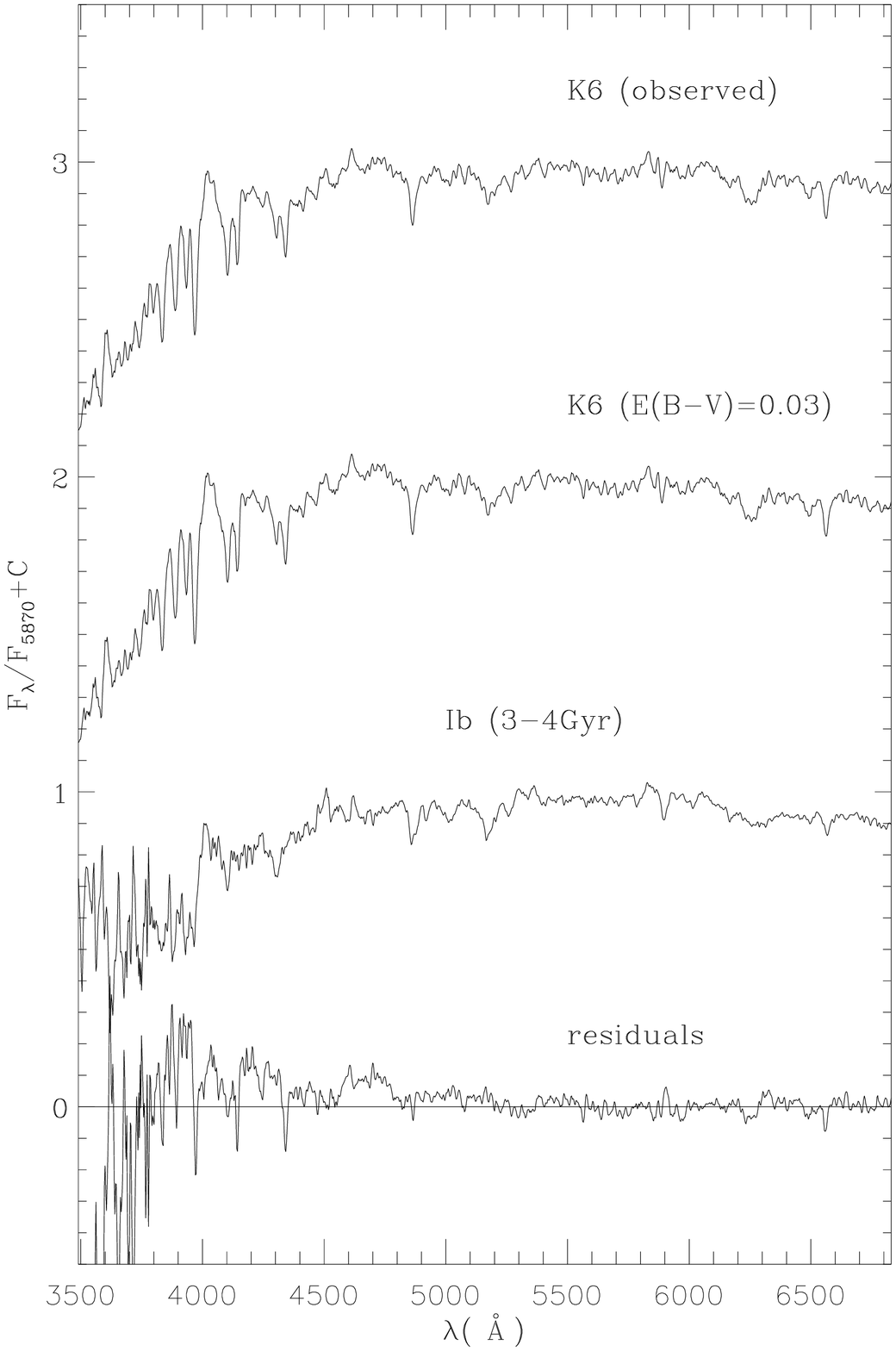}}
 \caption{Observed integrated spectrum of K\,6 (top), the spectrum corrected
for the adopted reddening $E(B-V)$ and the template spectrum which best
matches it (middle), and the residuals between both (bottom). See details 
in Sect. 3.3.4.}
\label{fig4}
\end{figure}

\begin{figure}
\resizebox{\hsize}{!}{\includegraphics{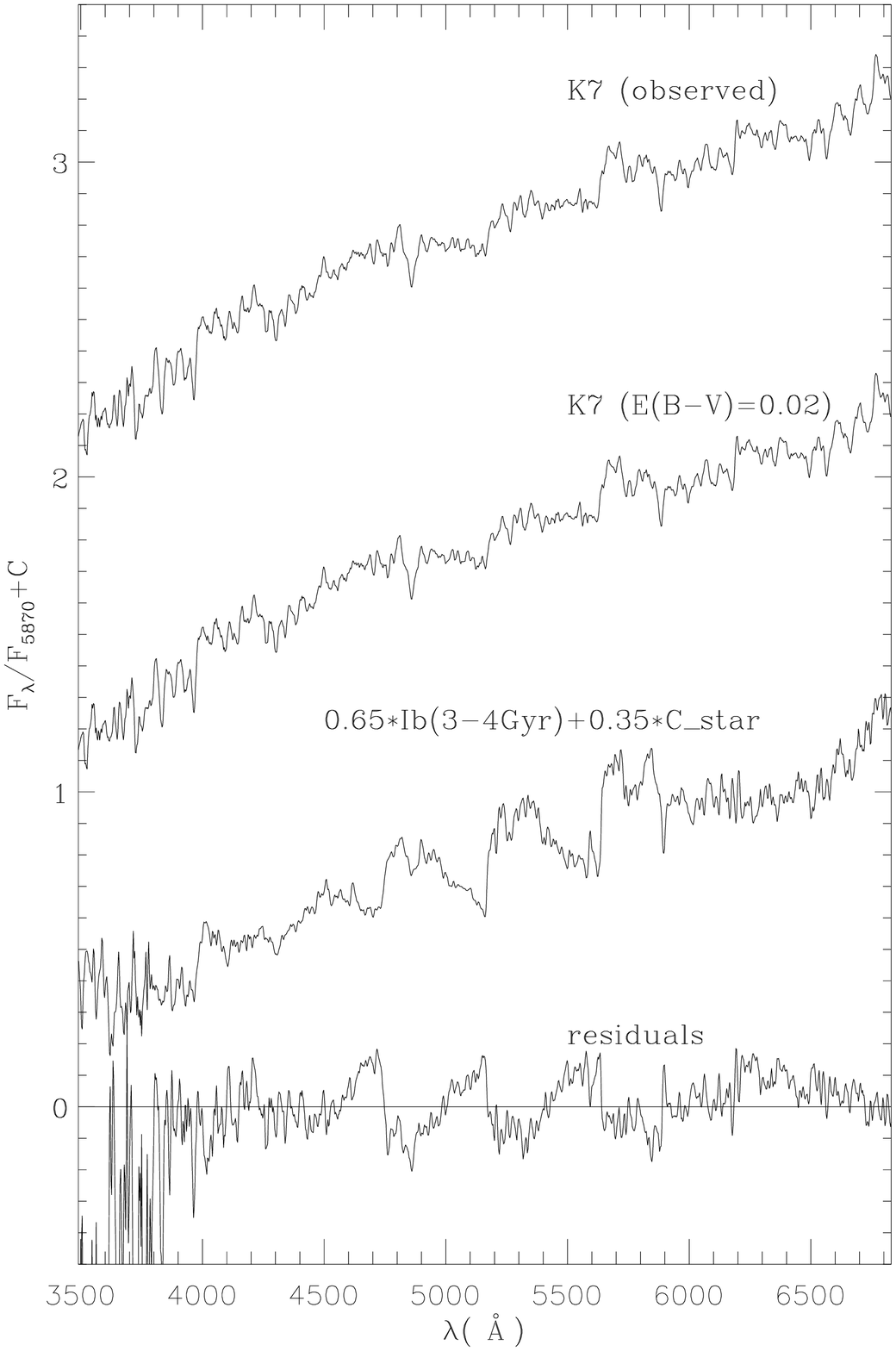}}
 \caption{Observed integrated spectrum of K\,7 (top), the spectrum corrected
for the adopted reddening $E(B-V)$ and the template spectrum which best
matches it (middle), and the residuals between both (bottom). See details 
in Sect. 3.3.5.}
\label{fig5}
\end{figure}

\begin{figure}
\resizebox{\hsize}{!}{\includegraphics{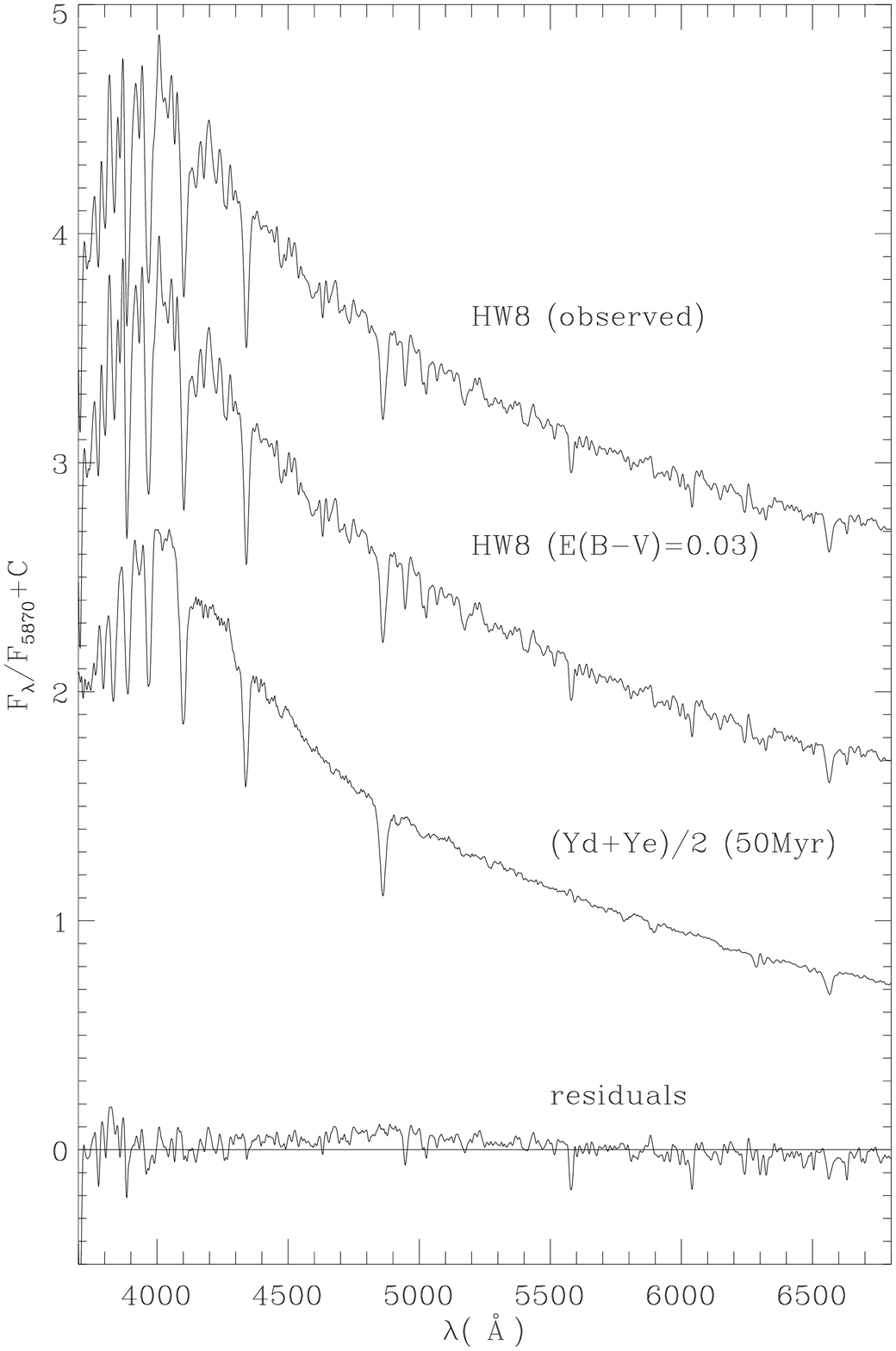}}
 \caption{Observed integrated spectrum of HW\,8 (top), the spectrum corrected
for the adopted reddening $E(B-V)$ and the template spectrum which best
matches it (middle), and the residuals between both (bottom).}
\label{fig6}
\end{figure}

\begin{figure}
\resizebox{\hsize}{!}{\includegraphics{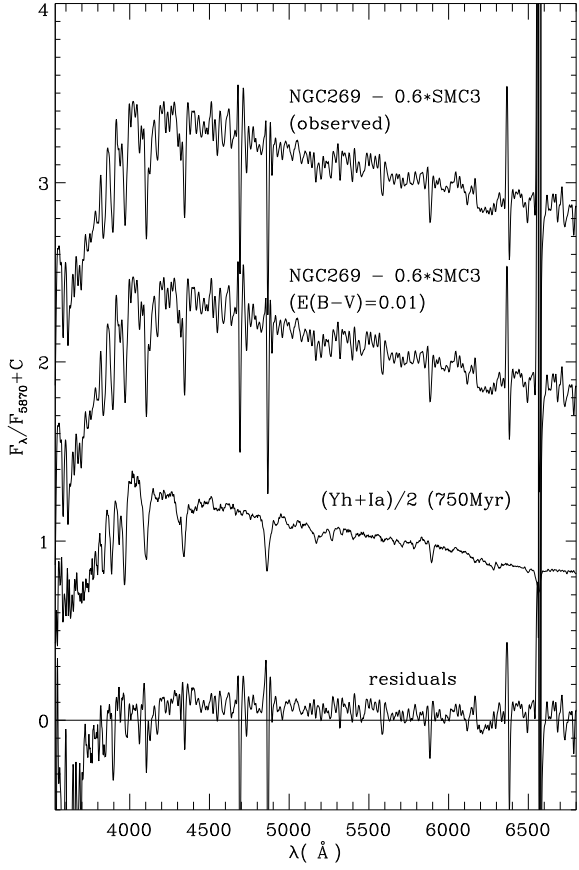}}
 \caption{Observed integrated spectrum of NGC\,269 with the symbiotic nova 
SMC\,3 (top) subtracted, the spectrum corrected
for the adopted reddening $E(B-V)$ and the template spectrum which best
matches it (middle), and the residuals between both (bottom). See details 
in Sect. 3.3.6.}
\label{fig7}
\end{figure}

\begin{figure}
\resizebox{\hsize}{!}{\includegraphics{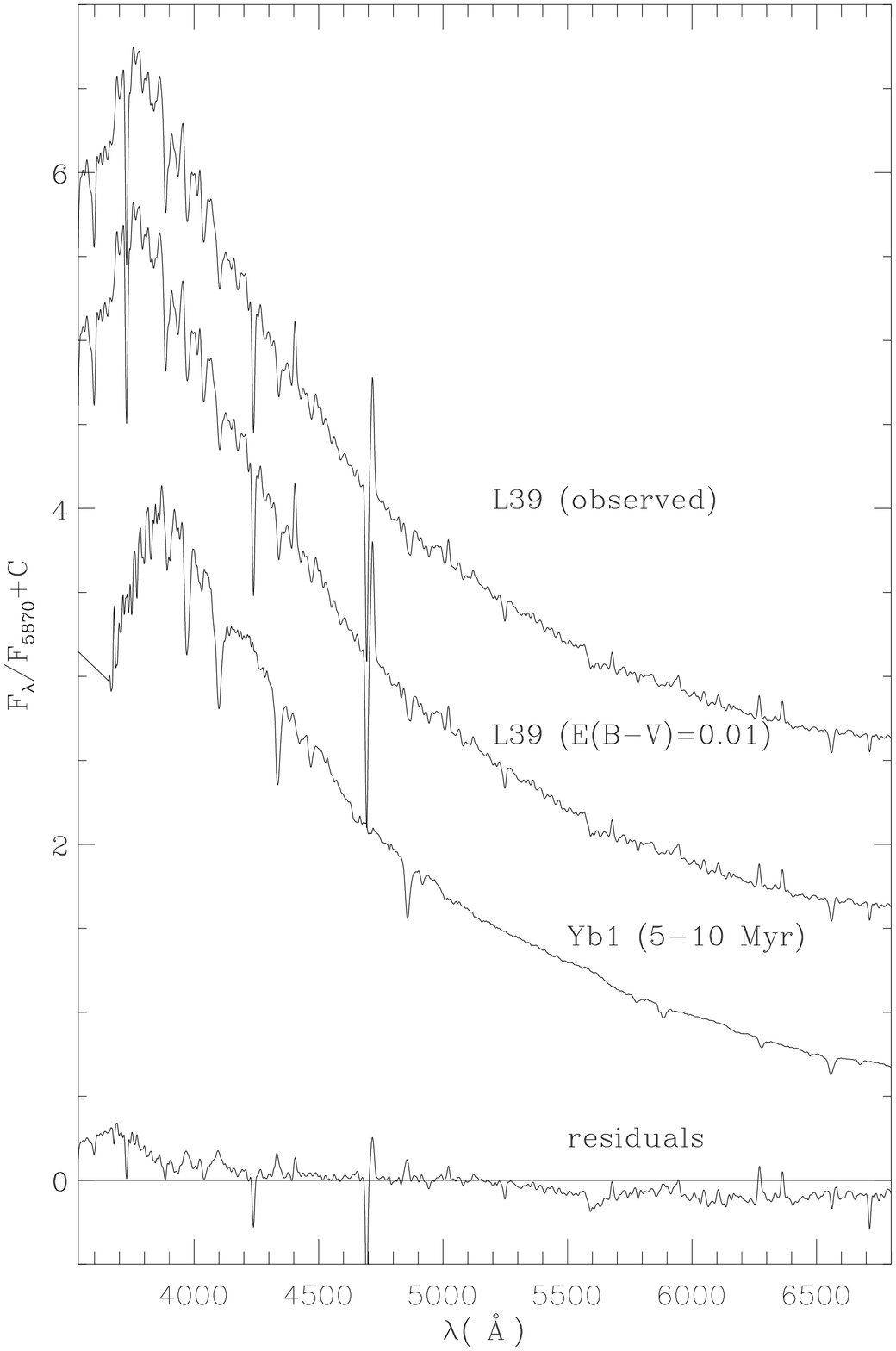}}
 \caption{Observed integrated spectrum of L\,39 (top), the spectrum 
corrected for the adopted reddening $E(B-V)$ and the template spectrum which 
best matches it (middle), and the residuals between both (bottom). See
details in Sect. 3.3.7.}
\label{fig8}
\end{figure}

\begin{figure}
\resizebox{\hsize}{!}{\includegraphics{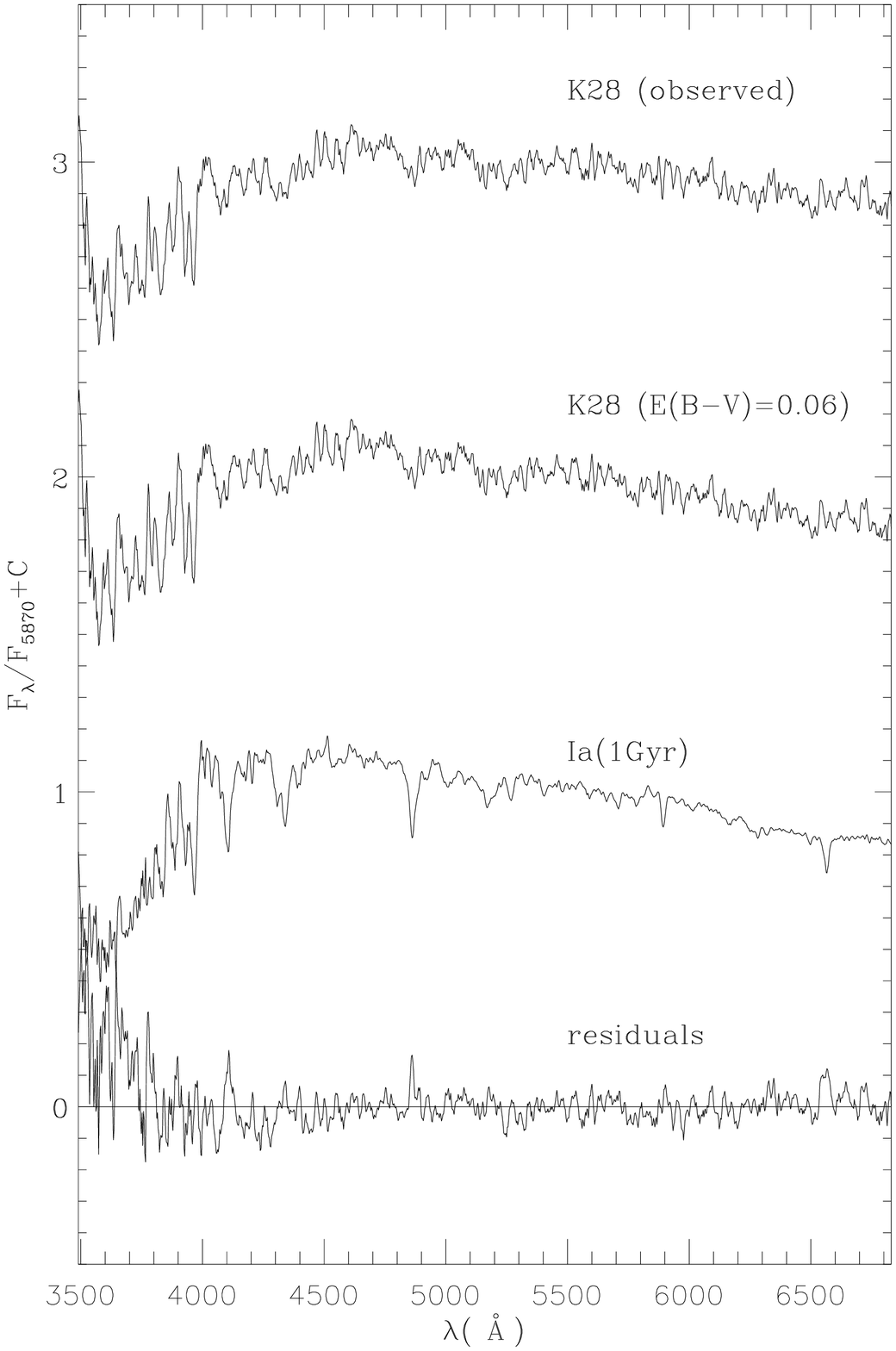}}
 \caption{Observed integrated spectrum of K\,28 (top), the spectrum corrected
for the adopted reddening $E(B-V)$ and the template spectrum which best
matches it (middle), and the residuals between both (bottom). See details 
in Sect. 3.3.8.}
\label{fig9}
\end{figure}

\begin{figure}
\resizebox{\hsize}{!}{\includegraphics{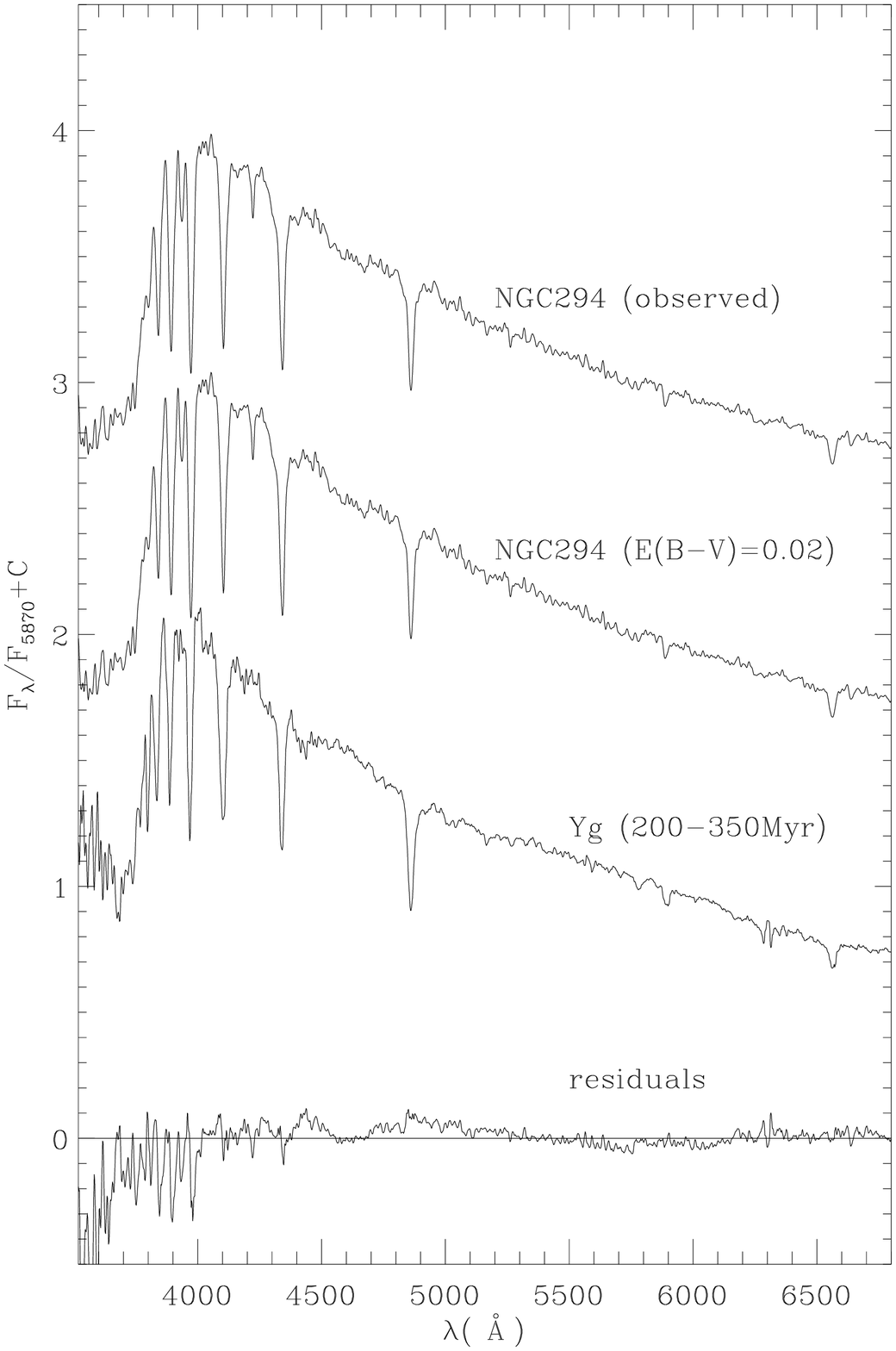}}
 \caption{Observed integrated spectrum of NGC\,294 (top), the spectrum 
corrected for the adopted reddening $E(B-V)$ and the template spectrum which 
best matches it (middle), and the residuals between both (bottom). See 
details in Sect. 3.3.9.}
\label{fig10}
\end{figure}

\begin{figure}
\resizebox{\hsize}{!}{\includegraphics{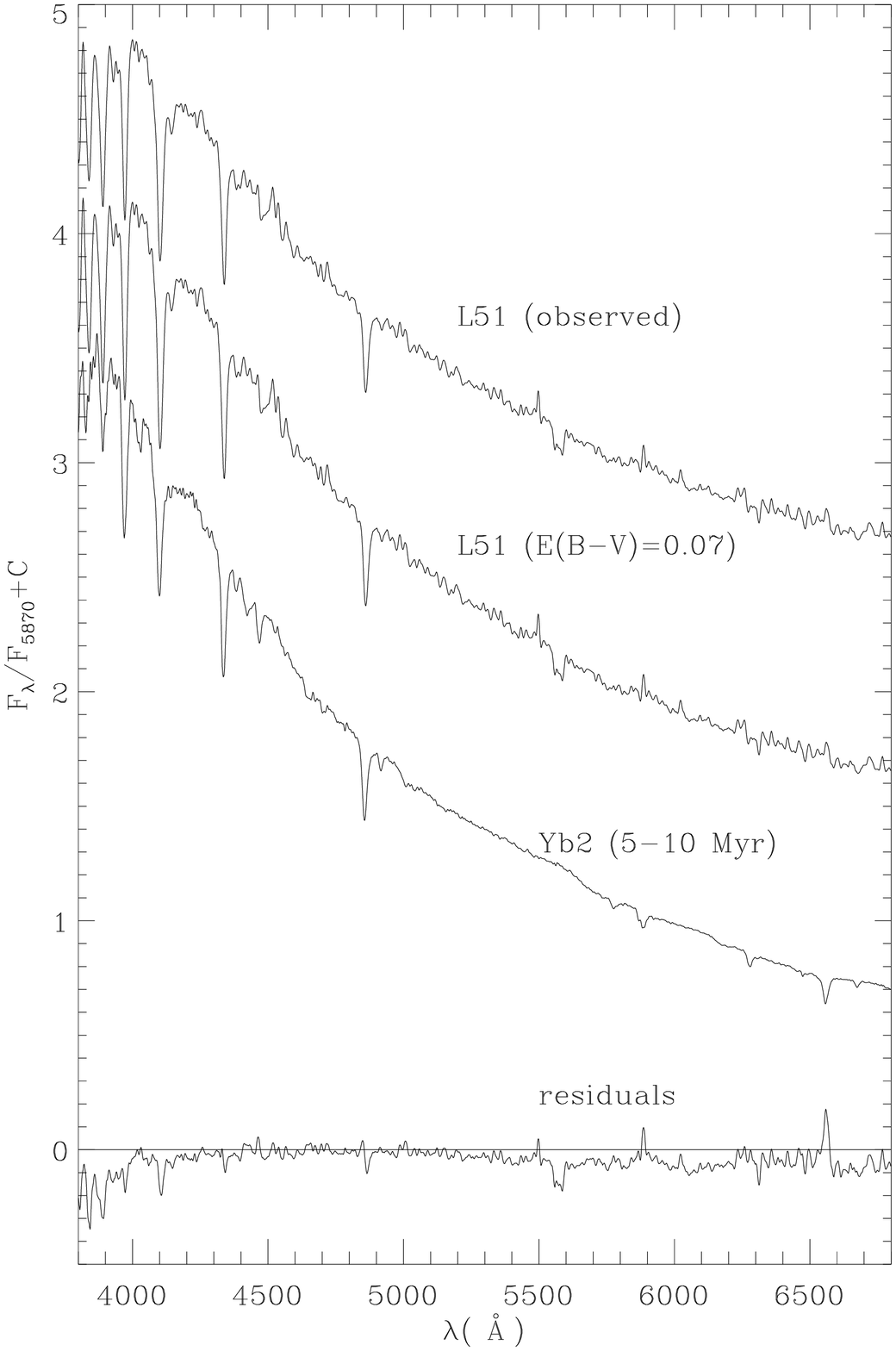}}
 \caption{Observed integrated spectrum of L\,51 (top), the spectrum corrected
for the adopted reddening $E(B-V)$ and the template spectrum which best
matches it (middle), and the residuals between both (bottom). See details 
in Sect. 3.3.10.}
\label{fig11}
\end{figure}

\begin{figure}
\resizebox{\hsize}{!}{\includegraphics{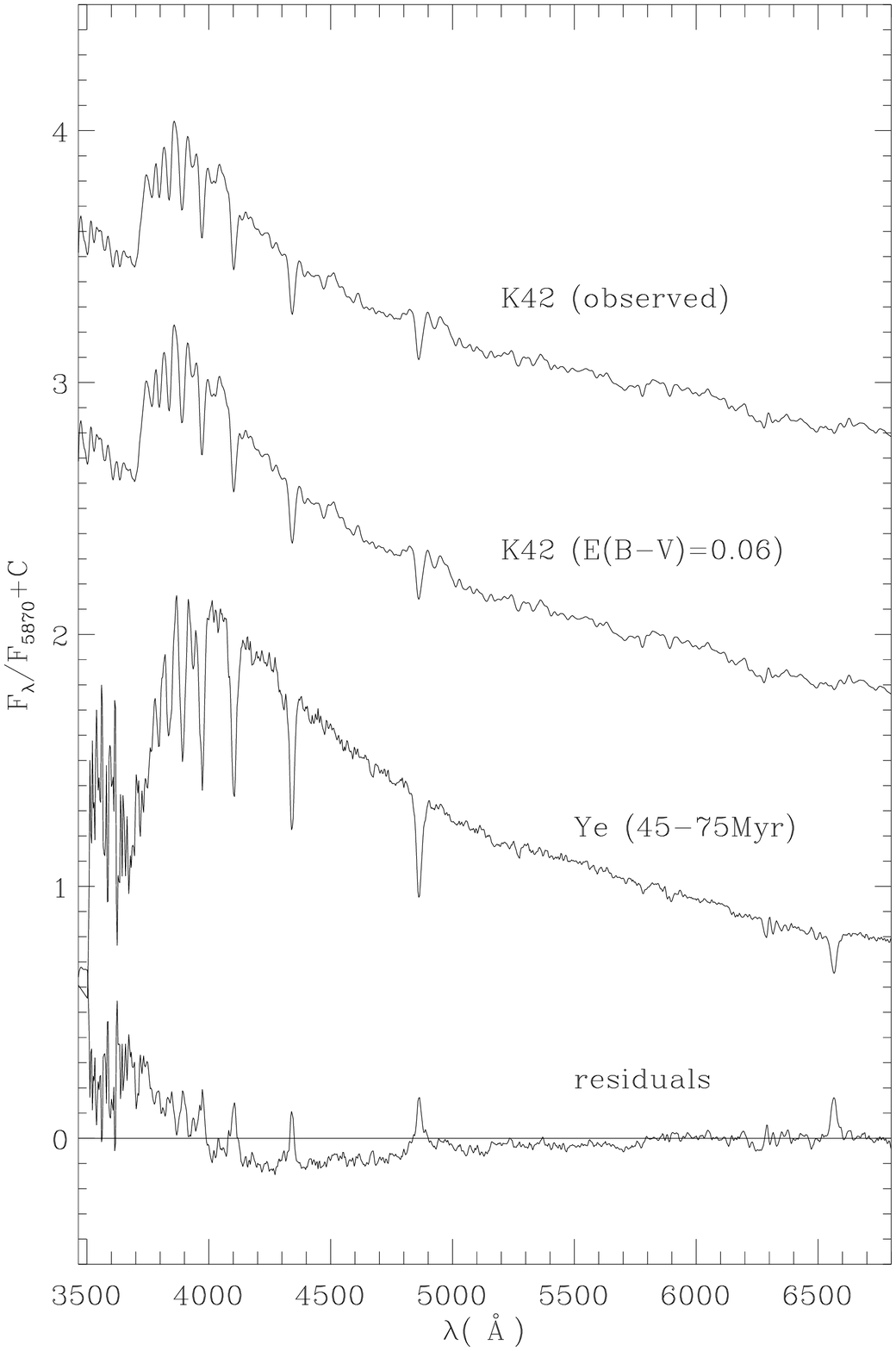}}
 \caption{Observed integrated spectrum of K\,42 (top), the spectrum corrected
for the adopted reddening $E(B-V)$ and the template spectrum which best
matches it (middle), and the residuals between both (bottom). See details 
in Sect. 3.3.11.}
\label{fig12}
\end{figure}

\begin{figure}
\resizebox{\hsize}{!}{\includegraphics{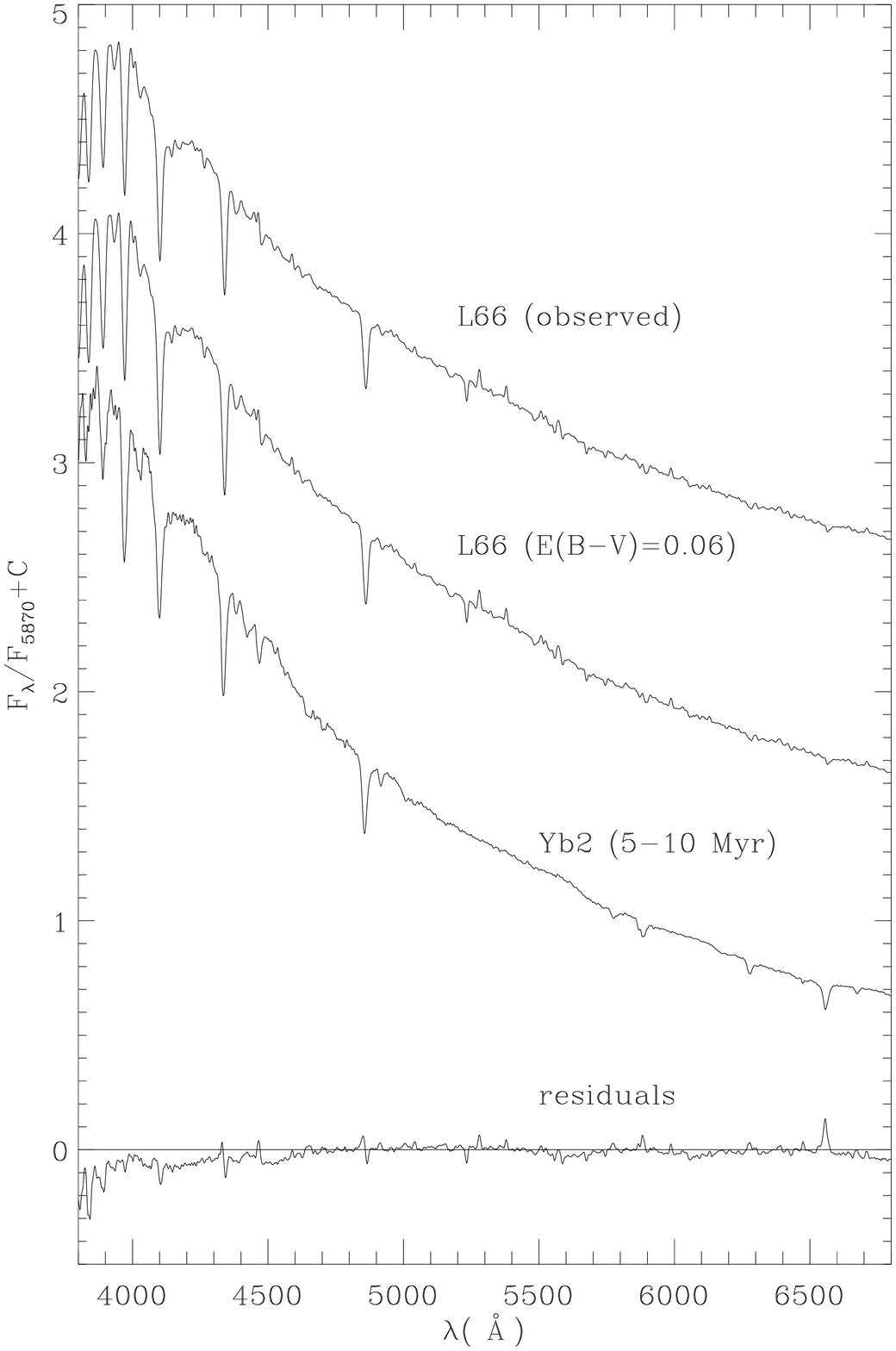}}
 \caption{Observed integrated spectrum of L\,66 (top), the spectrum corrected
for the adopted reddening $E(B-V)$ and the template spectrum which best
matches it (middle), and the residuals between both (bottom). See details 
in Sect. 3.3.12.}
\label{fig13}
\end{figure}

\begin{figure}
\resizebox{\hsize}{!}{\includegraphics{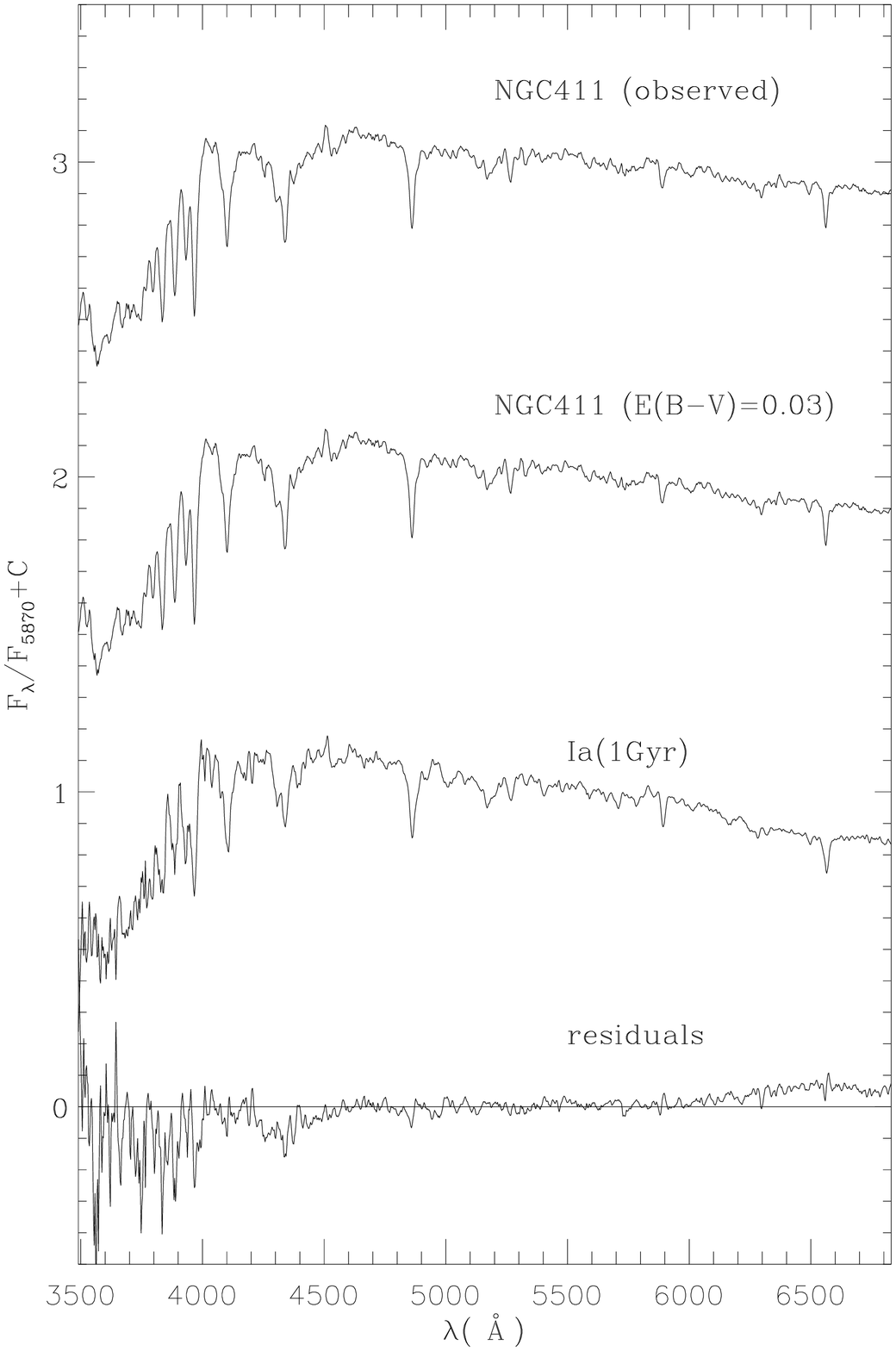}}
 \caption{Observed integrated spectrum of NGC\,411 (top), the spectrum 
corrected for the adopted reddening $E(B-V)$ and the template spectrum which 
best matches it (middle), and the residuals between both (bottom). See 
details in Sect. 3.3.13.}
\label{fig14}
\end{figure}

\begin{figure}
\resizebox{\hsize}{!}{\includegraphics{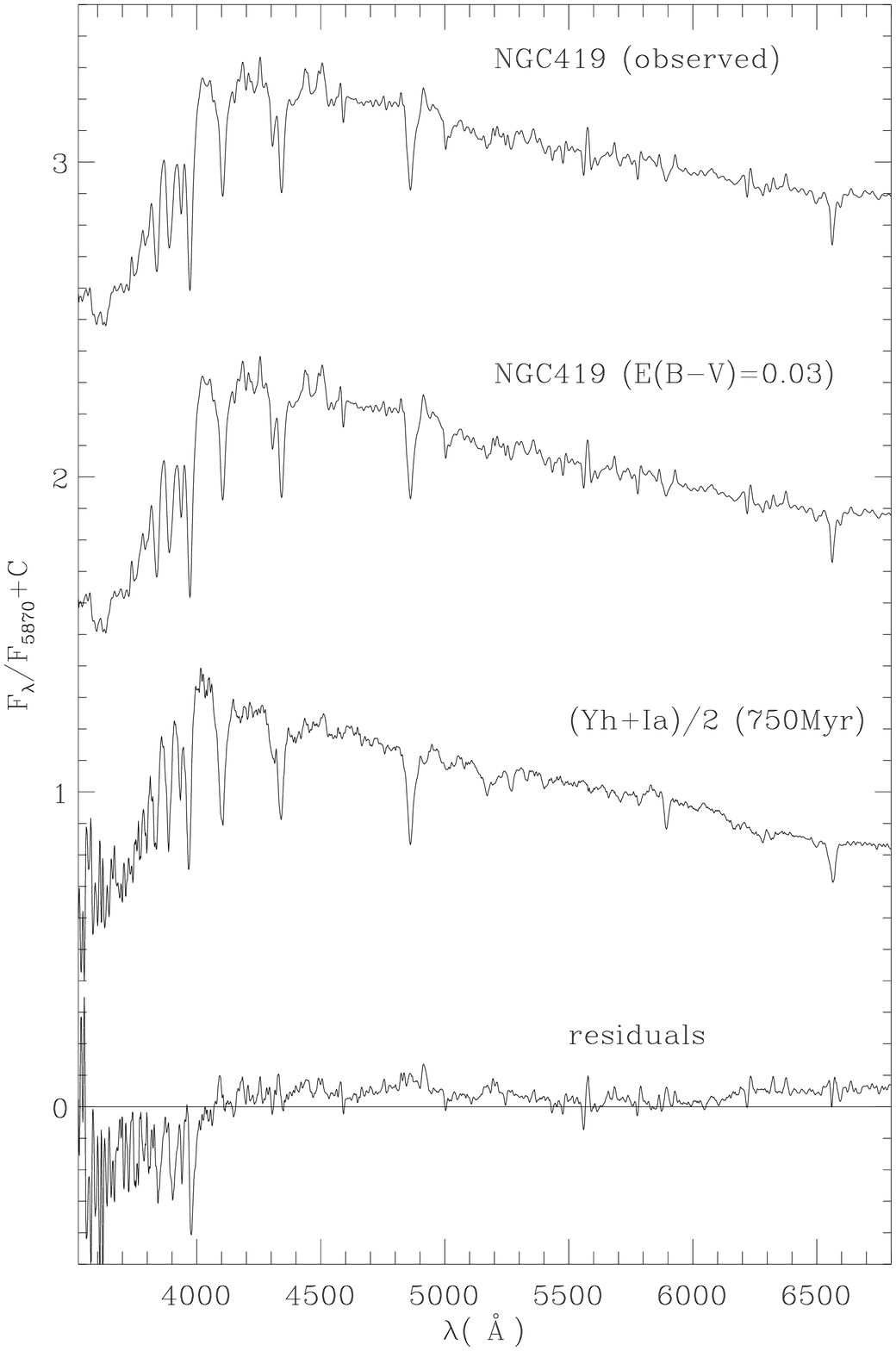}}
 \caption{Observed integrated spectrum of NGC\,419 (top), the spectrum 
corrected for the adopted reddening $E(B-V)$ and the template spectrum which 
best matches it (middle), and the residuals between both (bottom). See 
details in Sect. 3.3.14.}
\label{fig15}
\end{figure}

\begin{figure}
\resizebox{\hsize}{!}{\includegraphics{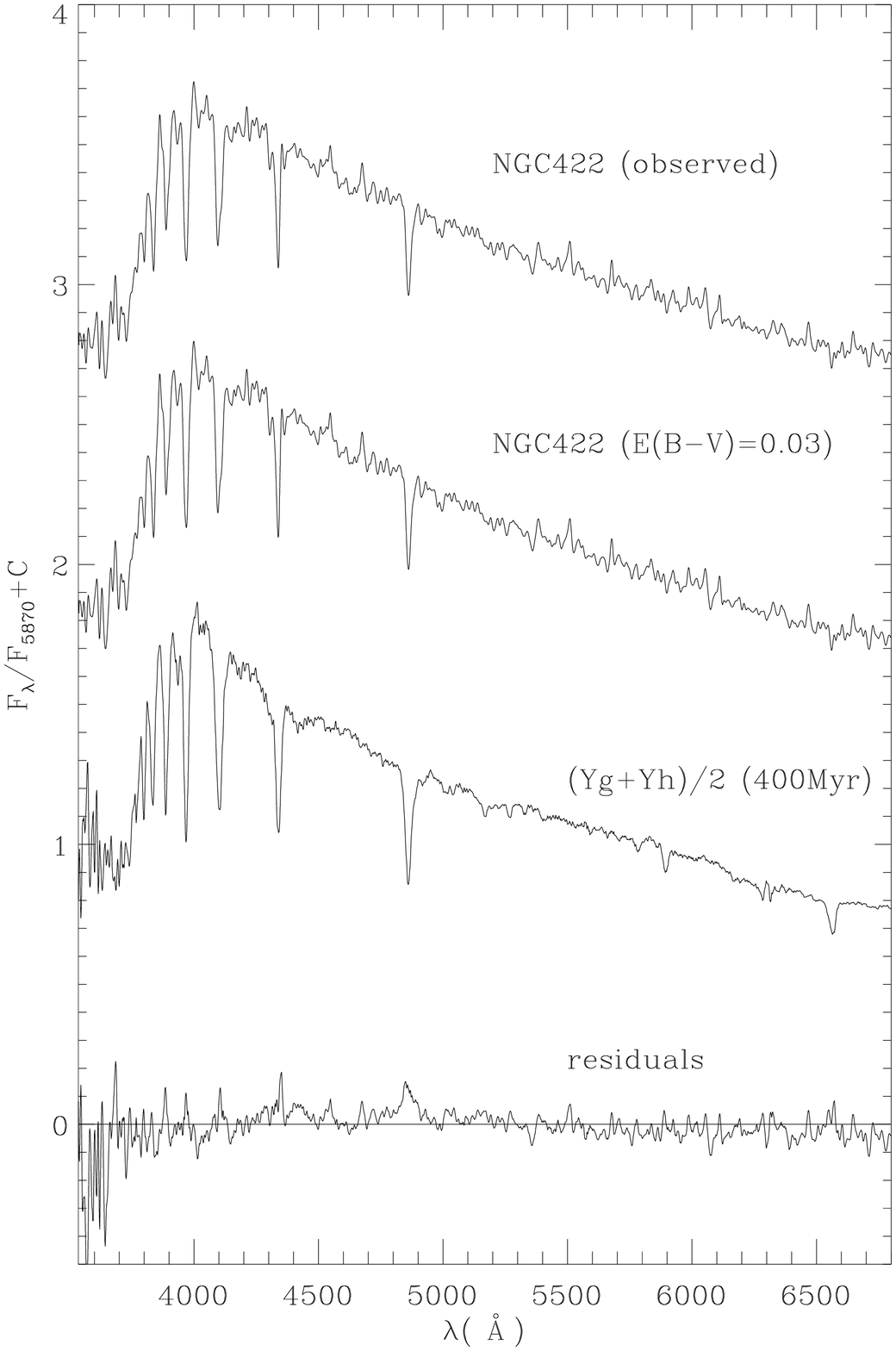}}
 \caption{Observed integrated spectrum of NGC\,422 (top), the spectrum 
corrected for the adopted reddening $E(B-V)$ and the template spectrum which 
best matches it (middle), and the residuals between both (bottom). See 
details in Sect. 3.3.15.}
\label{fig16}
\end{figure}

\begin{figure}
\resizebox{\hsize}{!}{\includegraphics{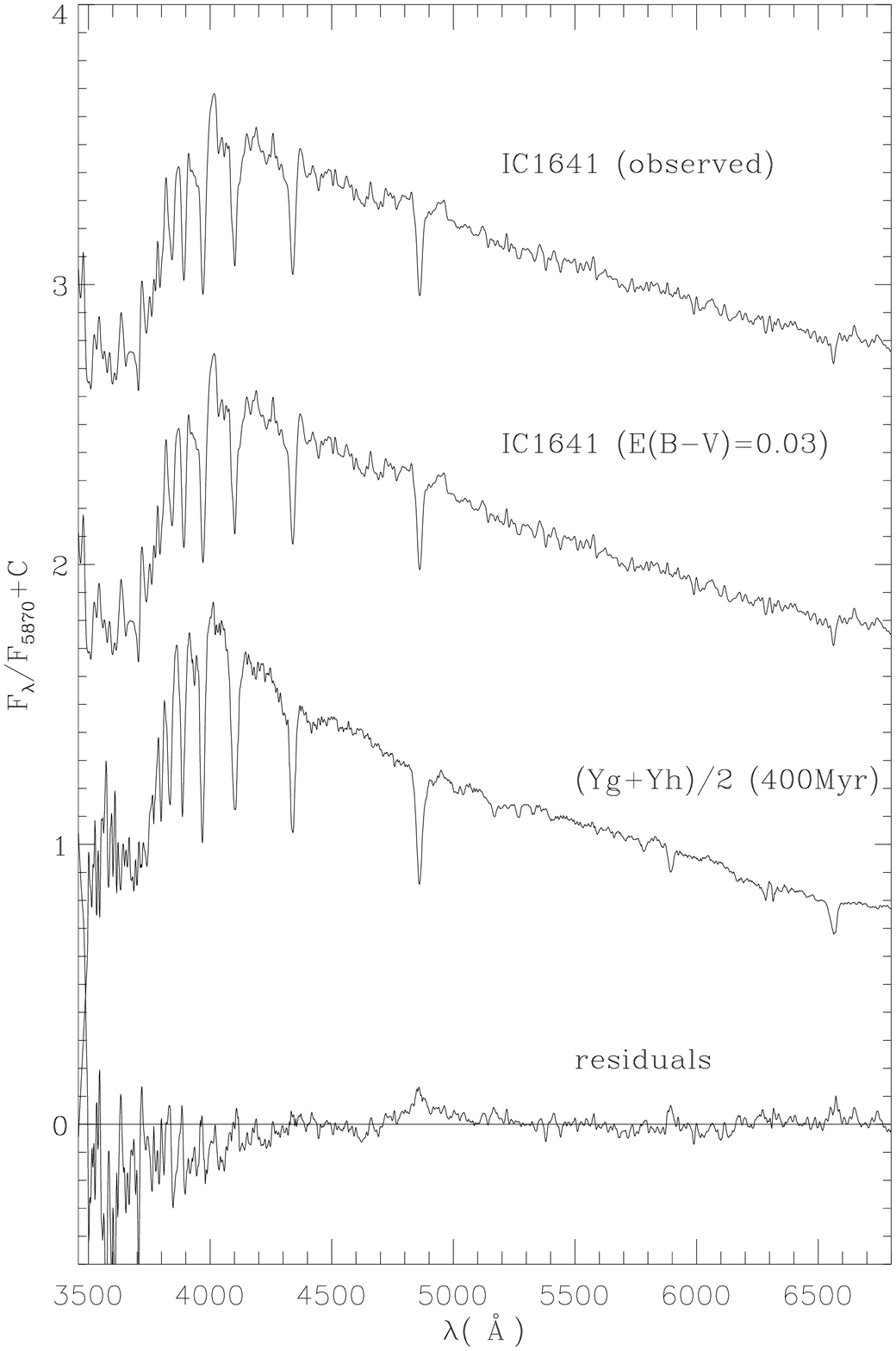}}
 \caption{Observed integrated spectrum of IC\,1641 (top), the spectrum 
corrected for the adopted reddening $E(B-V)$ and the template spectrum which 
best matches it (middle), and the residuals between both (bottom).}
\label{fig17}
\end{figure}

\begin{figure}
\resizebox{\hsize}{!}{\includegraphics{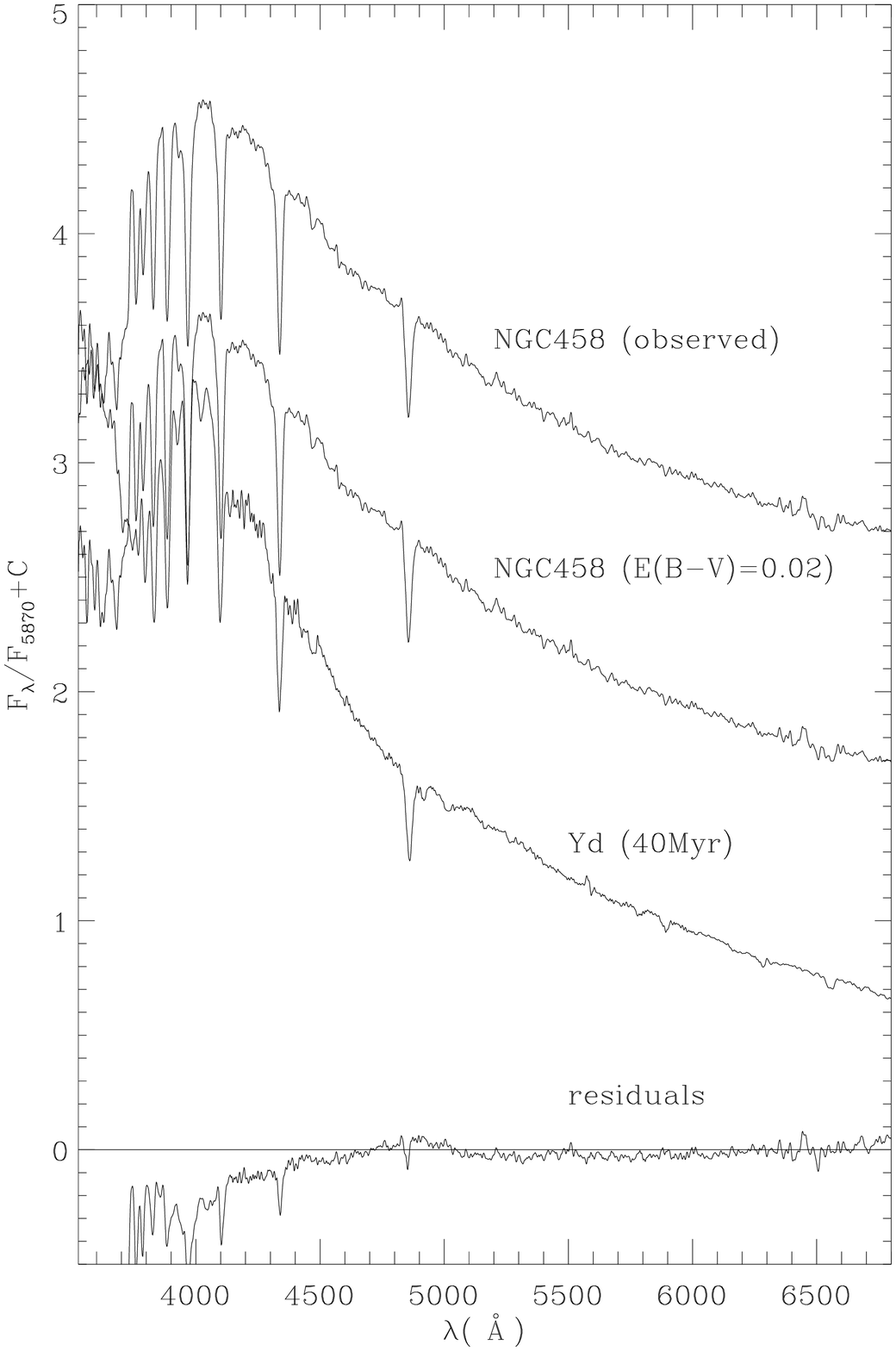}}
 \caption{Observed integrated spectrum of NGC\,458 (top), the spectrum 
corrected for the adopted reddening $E(B-V)$ and the template spectrum which 
best matches it (middle), and the residuals between both (bottom). See 
details in Sect. 3.3.16.}
\label{fig18}
\end{figure}

\clearpage

\begin{figure}
\resizebox{\hsize}{!}{\includegraphics{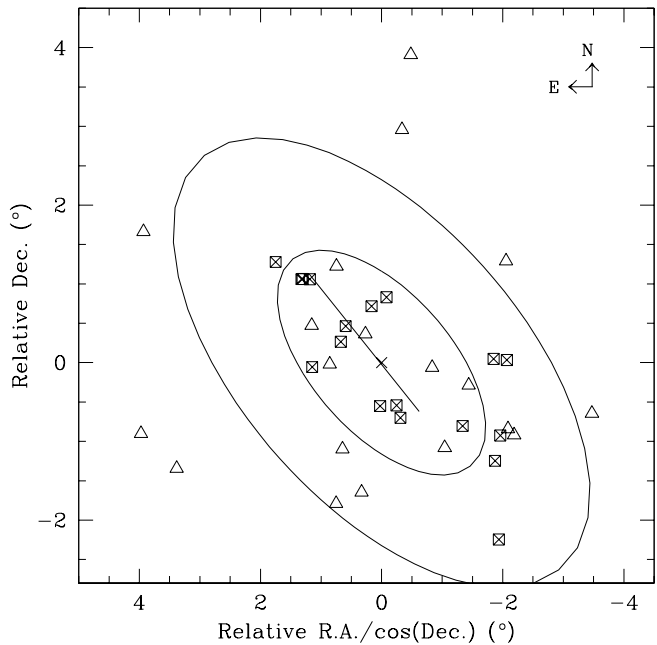}}
 \caption{Positions of the eighteen studied cluster fields (crossed boxes)
and of the nineteen additional clusters taken from \cite{petal02b} and 
\cite{petal05} (triangles) with relation to the SMC bar (straight line) and 
optical centre (cross). Two ellipses of semi-major axes of 2$\degr$ and 
4$\degr$ ($b/a$ = 1/2), aligned along the SMC Bar, are 
also drawn.}
\label{fig19}
\end{figure}

\begin{figure}
\resizebox{\hsize}{!}{\includegraphics{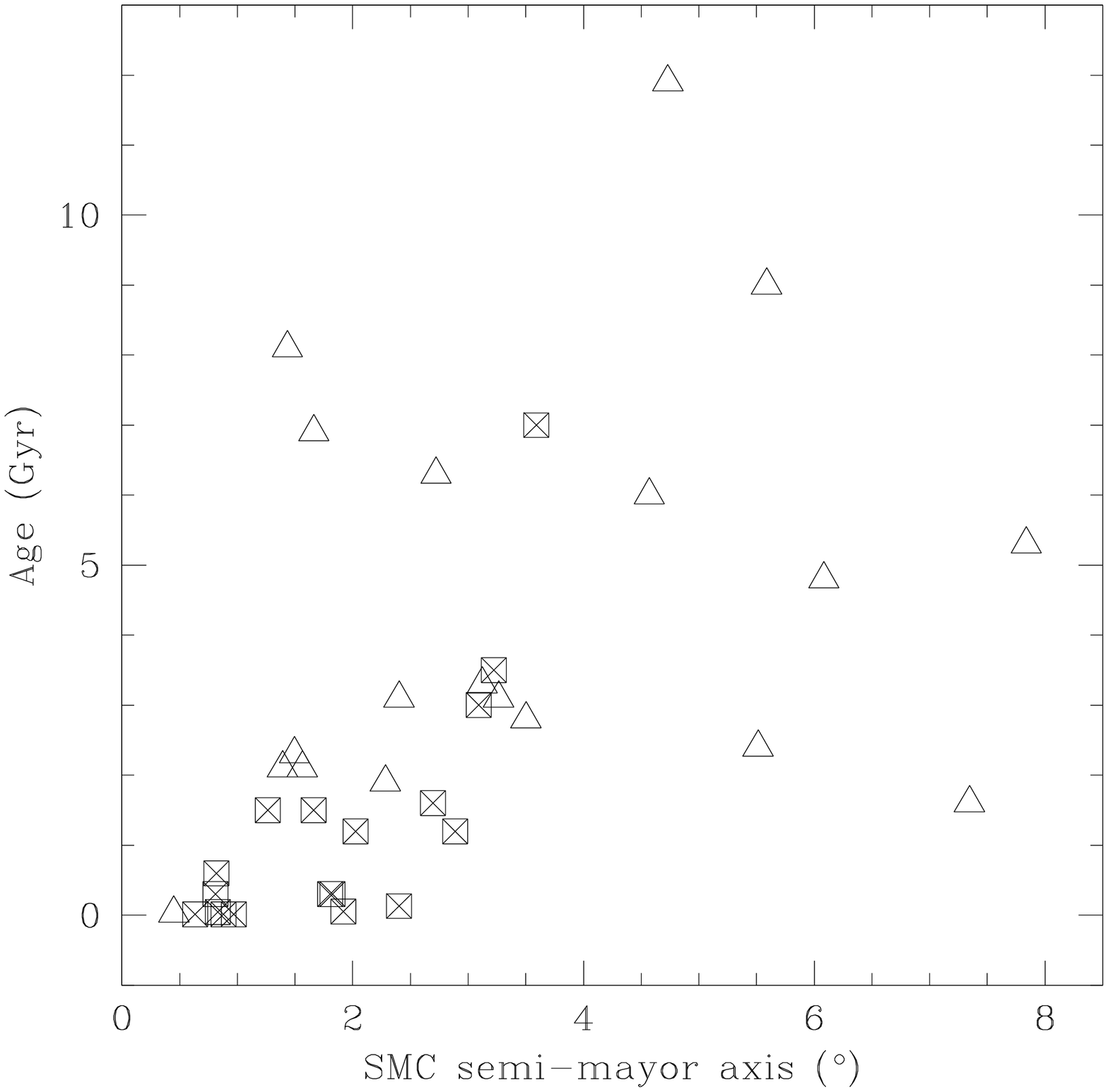}}
 \caption{Cluster ages versus semi-major axes of ellipses with $b/a$ = 
1/2, centred at the SMC optical centre, aligned along the SMC Bar, that
pass through the cluster positions. Symbols are as in Fig. 19.}
\label{fig20}
\end{figure}

\begin{figure}
\resizebox{\hsize}{!}{\includegraphics{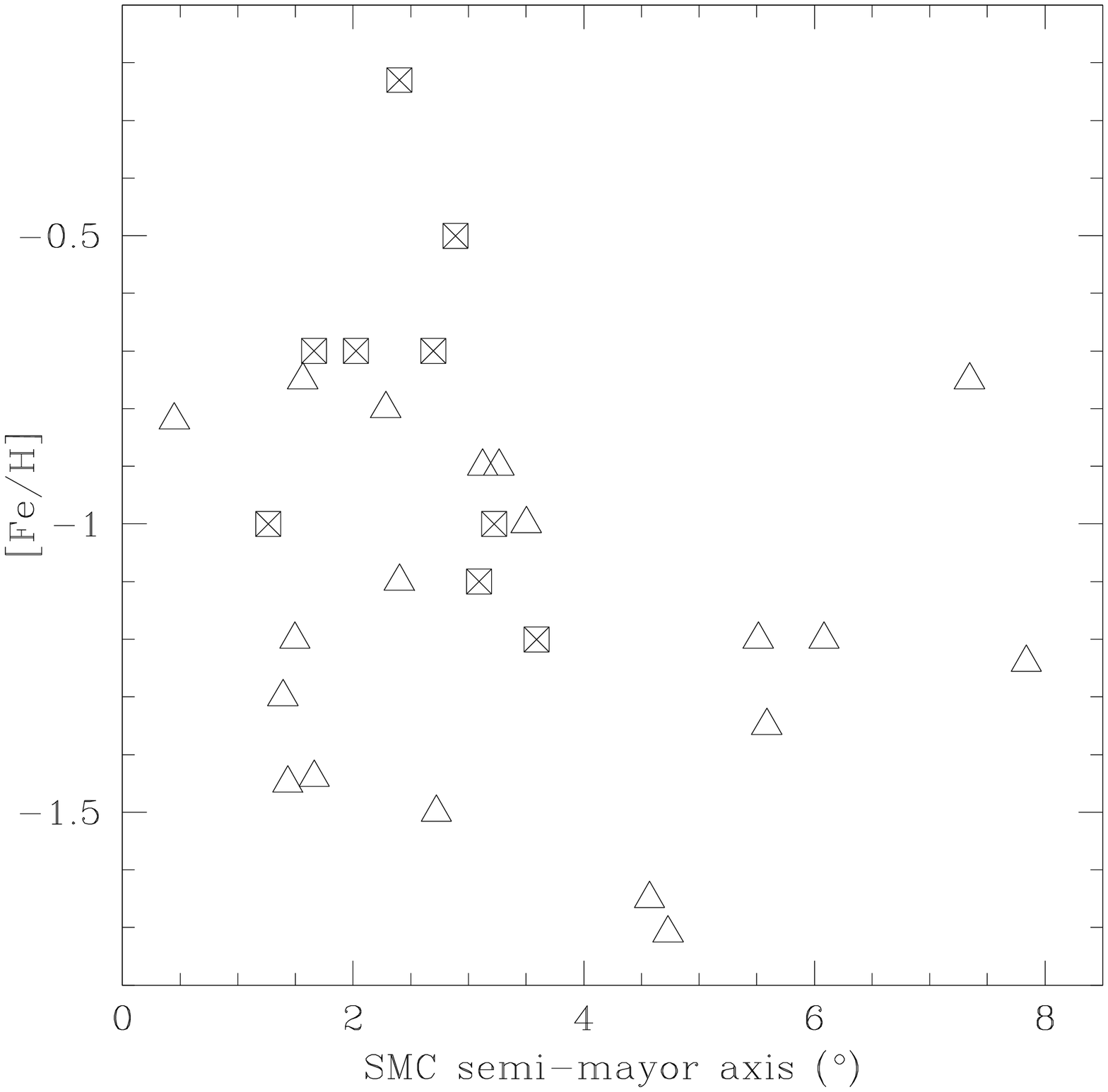}}
 \caption{Cluster metallicity versus semi-major axes of ellipses with $b/a$ 
= 1/2, centred at the SMC optical centre, aligned along the SMC Bar, that
pass through the cluster positions. Symbols are as in Fig. 19.}
\label{fig21}
\end{figure}

\clearpage

\begin{table}
\label{obs}
\caption{The sample clusters}
\begin{tabular}{lccccc}
\hline
Name$^a$  &  $\alpha_{\rm 2000}$ & $\delta_{\rm 2000}$ & D & Origin & S/N \\
      &  (h:m:s)  & ($\degr$:$\arcmin$:$\arcsec$) & ($\arcmin$) & &  \\
\hline

L\,5, ESO\,28-SC\,16                                   & 0:22:40 & -75:04:29 & 1.10 & CTIO   & 20\\
K\,5, L\,7, ESO\,28-SC\,18                             & 0:24:43 & -73:45:18 & 1.80 & CTIO   & 30\\
K\,3, L\,8, ESO\,28-SC\,19                             & 0:24:47 & -72:47:39 & 3.40 & CTIO   & 30\\
K\,6, L\,9, ESO\,28-SC\,20                             & 0:25:26 & -74:04:33 & 1.00 & CTIO   & 35\\
K\,7, L\,11, ESO\,28-SC\,22                            & 0:27:46 & -72:46:55 & 1.70 & CTIO   & 30\\
HW\,8                                                  & 0:33:46 & -73:37:59 & 1.70 & CASLEO & 20\\
NGC\,269, K\,26, L\,37, ESO\,29-SC\,16                 & 0:48:21 & -73:31:49 & 1.20 & CASLEO & 30\\
L\,39, SMC\_OGLE\,54                                   & 0:49:18 & -73:22:20 & 0.70 & CASLEO & 35\\
K\,28, L\,43, ESO\,51-SC\,4                            & 0:51:42 & -71:59:52 & 1.70 & CTIO   & 15\\
NGC\,294, L\,47, ESO\,29-SC\,22, SMC\_OGLE\,90         & 0:53:06 & -73:22:49 & 1.70 & CASLEO & 45\\
L\,51, ESO\,51-SC\,7                                   & 0:54:54 & -72:06:46 & 1.00 & CASLEO & 50\\
K\,42, L\,63, SMC\_OGLE\,124                           & 1:00:34 & -72:21:56 & 0.85 & CASLEO & 40\\
L\,66, SMC\_OGLE\,129                                  & 1:01:45 & -72:33:52 & 1.10 & CASLEO &100\\
NGC\,411, K\,60, L\,82, ESO\,51-SC\,19                 & 1:07:55 & -71:46:08 & 2.10 & CTIO   & 55\\
NGC\,419, K\,58, L\,85, ESO\,29-SC\,33, SMC\_OGLE\,159 & 1:08:19 & -72:53:03 & 2.80 & CASLEO & 70\\
NGC\,422, K\,62, L\,87, ESO\,51-SC\,22                 & 1:09:25 & -71:46:00 & 1.00 & CASLEO & 30\\
IC\,1641, HW\,62, ESO\,51-SC\,21                       & 1:09:39 & -71:46:07 & 0.75 & CASLEO & 45\\
NGC\,458, K\,69, L\,96, ESO\,51-SC\,26                 & 1:14:52 & -71:33:00 & 2.60 & CASLEO & 45\\

\hline
\end{tabular}

$^a$ Cluster identifications are from Kron (1956, K), 
Lindsay (1958, L), Hodge \& Wright (1974, HW), Lauberts 
(1982, ESO), Pietrzy\'nski et al. (1998, SMC\_OGLE).

\end{table}

\begin{table}
\label{ews}
\caption{Equivalent widths (\AA)}
\begin{tabular}{lrrrrrrrr}
\hline
Feature &   K\,Ca\,II  & H$_\delta$& G\,band (CH)& H$_\gamma$&H$_\beta$&Mg\,I & $S_h$ & $S_m$\\
Windows (\AA) &3908 &4082&4284       &4318  &4846&5156&&\\
              &-3952&-4124 &-4318    &-4364  &-4884&-5196&&\\
\hline
Cluster & &&&&&&   & \\
L\,5&           5.2  &7.3         &  5.1     &8.8         &7.2         & 4.0   & 23.3 & 14.3\\
K\,5&           6.9  &8.5         &  5.1     &9.7         &6.3         & 3.5   & 24.5 & 15.5\\
K\,3&           6.8  &5.4         &  6.2     &3.9         &4.5         & 3.4   & 13.9 & 16.5\\
K\,28&          9.1  &2.9         &  4.1     &3.4         &2.9         & 3.4   &  9.3 & 16.6\\
K\,6&           5.5  &8.2         &  4.7     &6.6         &4.6         & 4.0   & 19.4 & 14.3\\
K\,7&          10.4  &2.5         &  7.0     &4.7         &4.6         & 4.0   & 11.8 & 21.4\\
HW\,8&          4.4  &7.9         &  1.3     &6.3         &6.3         & 3.2   & 20.4 &  8.9\\
NGC\,269$^a$&   4.7 &11.9       & 3.3     & 9.6       & --         &   4.0  & --  & 11.9 \\
L\,39&          2.3  &3.1         &  1.3     &3.1         & 2.2        & 0.7   &  8.4 & 4.2 \\
K\,28&          9.1  &2.9         &  4.1     &3.4         &2.9         & 3.4   &  9.3 & 16.6\\
NGC\,294&       4.1  &10.3        &  2.1    &11.2         &7.0         & 1.7   & 28.5 & 7.8 \\
L\,51&          1.0  &7.6         &  0.6    & 4.3         &4.9         & 1.2   & 16.8 & 2.7 \\
K\,42&          1.2  &4.5         &  1.1    & 4.8         &4.2         & 0.5   & 13.4 & 2.8 \\
L\,66&          0.6  &5.8         & -0.3    & 2.7         &3.8         & 1.0   & 12.3 & 1.2 \\
NGC\,411&       6.5  &8.7         &  5.3     &8.9         &6.0         & 2.7   & 23.6 & 14.5\\
NGC\,419&       6.9  &8.2         &  3.5     &7.7         &7.1         & 1.9   & 22.9 &12.3 \\
NGC\,422&       2.1  &8.2         &  1.3     &6.2         &6.1         & 1.2   & 20.5& 4.6  \\
IC\,1641&       4.0  &9.6         &  2.3     &8.1         &6.2         & 1.5   & 23.9& 7.7  \\
NGC\,458&       1.9  &8.6         &  0.8     &8.3         &7.9         & 2.1   & 24.8& 4.8  \\

\hline
\end{tabular}

$^a$ The spectrum of the symbiotic nova SMC\,3 was subtracted from the
observed spectrum. 
\end{table}

\clearpage

\begin{table}
\label{param}
\caption{Cluster parameters}
\begin{tabular}{lccrccccr}
\hline
Cluster &$E(B-V)$ & $t_{\rm literature}$  &  Ref. &  $t_{\rm m}$ & method  &  $t_{\rm adopted}$  &  [Fe/H] &  Ref. \\
        &       &(Gyr)   &       & (Gyr)&         &    (Gyr)            &           &       \\

\hline

L\,5     & 0.03 & 4.1           & 1     & 0.8   & $S_h$,$S_m$ - template & 3.0$\pm$1.5     &  -1.1$\pm$0.2 & 1,9 \\   
K\,5     & 0.02 & 2.0           & 1     & 0.8   & $S_h$,$S_m$ - template & 1.2$\pm$0.5     &  -0.5$\pm$0.2 & 1,9 \\ 
K\,3     & 0.02 & 7.0$\pm$1.0   & 2,4   & 7.0   & $S_h$,$S_m$ + template & 7.0$\pm$1.0     & -1.20$\pm$0.2 & 2,9 \\ 
K\,6     & 0.03 & 1.3           & 7     & 2.0   & $S_h$,$S_m$ + template & 1.6$\pm$0.4     & -0.7          &   9 \\ 
K\,7     & 0.02 & 3.5           & 3     & 4.0   & $S_h$,$S_m$ + template & 3.5$\pm$0.5     & -1.0          &   3 \\ 
HW\,8    & 0.03 &               &       & 0.05  & $S_h$,$S_m$ - template & 0.05$\pm$0.02   &               &     \\
NGC\,269 & 0.01 &               &       & 0.6   & $S_m$ - template       & 0.6$\pm$0.2     &               &     \\
L\,39    & 0.01 &               &       & 0.015 & $S_h$,$S_m$ - template & 0.015$\pm$0.010 &               &     \\
K\,28    & 0.06 &  2.1          & 2     & 1.0   &    template            & 1.5$\pm$0.6     & -1.0$\pm$0.2  & 2,9 \\
NGC\,294 & 0.02 &               &       & 0.3   & $S_h$,$S_m$ - template & 0.3$\pm$0.1     &               &     \\
L\,51    & 0.07 &               &       & 0.015 & $S_h$,$S_m$ - template & 0.015$\pm$0.010 &               &     \\
K\,42    & 0.06 &               &       & 0.045 & $S_h$,$S_m$ - template & 0.045$\pm$0.015 &               &     \\
L\,66    & 0.06 &               &       & 0.015 & $S_h$,$S_m$ - template & 0.015$\pm$0.010 &               &     \\
NGC\,411 & 0.03 & 1.5$\pm$0.2   & 2,6,8 & 1.0   & $S_h$,$S_m$ - template & 1.5$\pm$0.3     & -0.7$\pm$0.2  &2,6,8\\ 
NGC\,419 & 0.03 & 1.6$\pm$0.4   & 2,4   & 0.8   & $S_h$,$S_m$ - template & 1.2$\pm$0.4     & -0.7          & 2   \\
NGC\,422 & 0.03 &               &       & 0.3   & $S_h$,$S_m$ - template & 0.3$\pm$0.1     &               &     \\
IC\,1641 & 0.03 &               &       & 0.3   & $S_h$,$S_m$ - template & 0.3$\pm$0.1     &               &     \\
NGC\,458 & 0.02 & 0.17$\pm$0.03 & 2,5   & 0.05  & $S_h$,$S_m$ - template & 0.13$\pm$0.06   & -0.23         & 2   \\ 

\hline
\end{tabular}

References: (1) Piatti et al. (2005); (2) Piatti et al. (2002b); 
(3) Mould et al. (1992); (4) Rich et al. (2000); (5) 
Alcaino et al. (2003); (6) Alves \& Sarajedini (1999); (7)
Matteucci et al. (2002); (8) Leonardi \& Rose (2003); (9) this work.

\end{table}

\bibliographystyle{apj}

\begin{thebibliography}{}

\bibitem[Abia \& Isern(2000)]{ai00}
Abia, C., \& Isern, J. 2000, ApJ, 536, 438

\bibitem[Ahumada et al.(2002)]{aetal02}
Ahumada, A.V., Clari\'a, J.J., Bica, E., \& Dutra, C.M, 2002, A\&A, 393, 855

\bibitem[Alcaino et al.(2003)]{aetal03}
Alcaino, G., Alvarado, F., Borissova, J., \& Kurtev, R. 2003, A\&A, 400, 917

\bibitem[Alves \& Sarajedini(1999)]{as99}
Alves, D.R., \& Sarajedini, A. 1999, ApJ, 511, 225

\bibitem[Barnbaum et al.(1996)]{bsk96}
Barnbaum, C., Stone, R.P.S., \& Keenan, P.C. 1996, ApJS, 105, 419

\bibitem[Bellazzini et al.(2004)]{betal04}
Bellazzini, M., Ibata, R., Monaco, L., Martin, N., Irwin, M.J., \& Lewis, 
G.F., 2005, MNRAS, in press

\bibitem[Bekki et al.(2004)]{bekki04}
Bekki, K., Couch, W.J., Beasley, M.A., Forbes, D.A., Chiba, M., \& Da Costa, 
G.S., 2004, ApJ, 610, L93

\bibitem[Bica(1988)]{b88}
Bica, E. 1988, A\&A, 195, 76

\bibitem[Bica \& Alloin(1986)]{ba86}
Bica, E., \& Alloin, D. 1986, A\&A, 162, 21

\bibitem[Bica et al.(1986)]{bdp86}
Bica E., Dottori H., \& Pastoriza M. 1986, A\&A 156, 261

\bibitem[Bica \& Dutra(2000)]{bd00}
Bica, E., \& Dutra, C.M. 2000, AJ, 119, 1214

\bibitem[Bica \& Schmitt(1995)]{bs95}
Bica, E., \& Schmitt, H.R., 1995, ApJS, 101, 41

\bibitem[Brocato et al.(2001)]{bdm01}
Brocato, E., Di Carlo, E., \& Menna, G. 2001, A\&A, 374, 523

\bibitem[Burstein \& Heiles(1982)]{bh82}
Burstein, D., \& Heiles, C. 1982, AJ, 87, 1165

\bibitem[Cioni et al(2000)]{cetal00}
Cioni, M.R., van der Marel, R.P., Loup, C., \& Habing, H.J. 2000, 
A\&A, 359, 601

\bibitem[Crowl et al.(2001)]{cetal01}
Crowl, H. H., Sarajedini, A., Piatti, A.E., Geisler, D., Bica, E., 
Clari\'a, J.J., \& Santos Jr., J.F.C. 2001, AJ, 122, 220

\bibitem[Da Costa(1991)]{d91}
Da Costa, G.S., 1991, IAUS, 148, 183

\bibitem[Da Costa \& Armandroff(1995)]{da95}
Da Costa, G.S., \& Armandroff, T.E., 1995, AJ, 109, 2533


\bibitem[Da Costa \& Mould(1986)]{dm86}
Da Costa, G.S., \& Mould, J.R. 1986, ApJ, 305, 214

\bibitem[de Oliveira et al.(2000)]{odbd00}
de Oliveira, M.R., Dutra, C.M., Bica, E., \& Dottori, H. 2000,
A\&AS, 146, 57

\bibitem[Dolphin et al.(2001)]{detal01}
Dolphin, A.E., Walker, A.R., Hodge, P.W., Mateo, M., Olszewski, E.W., 
Schommer, R.A., \& Suntzeff, N.B. 2001, ApJ, 562, 303

\bibitem[Durand et al.(1984)]{dhm84}
Durand, D., Hardy, E., \& Melnick, J. 1984, ApJ, 283, 552

\bibitem[Dutra et al.(2001)]{duetal01}
Dutra, C.M., Bica, E., Clari\'a, J.J., Piatti, A.E., \& Ahumada, A.V. 
2001, A\&A, 371, 895

\bibitem[Dutra et al.(1999)]{dbc99}
Dutra, C.M., Bica, E., Clari\'a, J.J., \& Piatti, A.E. 1999, MNRAS, 305, 373

\bibitem[Elson \& Fall(1988)]{ef88}
Elson, R.A.W., \& Fall, S.M. 1988, AJ, 96, 1383

\bibitem[Girardi et al.(2002)]{gbb02}
Girardi, L., Bertelli, G., Bressan, A., et al. 2002, A\&A, 391, 195

\bibitem[Gonz\'alez et al.(2004)]{glb04}
Gonz\'alez, R.A., Liu, M.C., \& Bruzual, G.A. 2004, ApJ, 611, 270

\bibitem[Harris \& Zaritsky(2004)]{hz04}
Harris, J., \& Zaritsky, D. 2004, AJ, 127, 1531

\bibitem[Hodge(1988)]{h88}
Hodge, P. 1988, PASP, 100, 1051

\bibitem[Hodge(1989)]{h89}
Hodge, P. 1989, ARA\&A, 27, 139

\bibitem[Hodge \& Wright(1974)]{hw74}
Hodge, P.W., \& Wright, F.W. 1974, AJ, 79, 858 

\bibitem[Kahabka(2004)]{k04}
Kahabka, P. 2004, A\&A, 416, 57

\bibitem[Kron(1956)]{k56}
Kron, G.E. 1956, PASP, 68, 125

\bibitem[Lauberts(1982)]{l82}
Lauberts, A. 1982, The ESO/Uppsala Survey of the ESO (B) Atlas, European 
Southern Observatory, Garching bei Munchen

\bibitem[Leonardi \& Rose(2003)]{lr03}
Leonardi, A.J., \& Rose, J.A. 2003, AJ, 126, 1811

\bibitem[Lin et al.(1995)]{ljk95}
Lin, D.N.C., Jones, B.F., \& Klemola, A.R. 1995, ApJ, 439, 652

\bibitem[Lindsay(1958)]{l58}
Lindsay, E.M. 1958, MNRAS, 118, 172

\bibitem[Matteucci et al.(2002)]{mrb02}
Matteucci, A., Ripepi, V., Brocato, E., \& Castellani, V. 2002, A\&A, 387, 861

\bibitem[Mighell et al.(1998)]{msf98}
Mighell, K.J., Sarajedini, A., \& French, R.S. 1998, AJ, 116, 2395

\bibitem[Mould et al.(1992)]{mjd82}
Mould, J.R., Jensen, J.B., \& Da Costa, G.S. 1992, ApJS, 82, 489

\bibitem[Munari \& Zwitter(2002)]{mz02}
Munari, U., \& Zwitter, T. 2002, A\&A, 383, 188

\bibitem[Pagel \& Tautvaisiene(1999)]{pt99}
Pagel, B.E.J., \& Tautvaisiene, G. 1999, Ap\&SS, 265, 461

\bibitem[Papenhausen \& Schommer(1988)]{ps88}
Papenhausen, P., \& Schommer, R.A., 1988, in IAU Symp. 
126, eds. J. Grindlay, A. G. Davis Philip. Dordrecht: 
Kluwer Academic Publishers, p.565

\bibitem[Piatti et al.(2002a)]{pbc02}
Piatti, A. E., Bica, E., Clari\'a, J.J., Santos Jr., J.F.C., 
\& Ahumada A.V. 2002a, MNRAS, 335, 233

\bibitem[Piatti et al.(2001)]{psc01}
Piatti A.E., Santos Jr., J.F.C., Clari\'a, J.J., et al. 2001, MNRAS, 325, 792

\bibitem[Piatti et al.(2002b)]{petal02b}
Piatti, A.E., Sarajedini, A., Geisler, D., Bica, E., \& Clari\'a, J.J. 2002b, 
MNRAS, 329, 556

\bibitem[Piatti et al.(2005)]{petal05}
Piatti, A.E., Sarajedini, A., Geisler, D., Seguel, J., \& Clark, D. 2005, 
MNRAS, 358, 1215

\bibitem[Pietrzynski et al.(1998)]{puk98}
Pietrzynski, G., Udalski, A., Kubiat, M., Szyma\'nski, M., 
Wo\'zniak, P., \& Zebru\'n, K. 1998, Acta Astron., 48, 175

\bibitem[Pietrzynski \& Udalski(1999)]{pu99}
Pietrzynski, G., \& Udalski, A. 1999, AcA, 49, 157

\bibitem[Rabin(1982)]{r82}
Rabin, D. 1982, ApJ, 261, 85

\bibitem[Rich et al.(1984)]{rcm84}
Rich, R. M., Da Costa, G. S., \& Mould, J. R. 1984, ApJ, 286, 517

\bibitem[Rich et al.(2000)]{retal00}
Rich, R.M., Shara, M., Fall, S.M., \& Zurek, D. 2000, AJ, 119, 197

\bibitem[Santos et al.(2002)]{sab02}
Santos Jr., J.F.C., Alloin, D., Bica, E., \& Bonatto, C. 2002, in IAU Symp. 
207, eds. D. Geisler, E. K. Grebel, and D. Minniti. San Francisco: 
Astronomical Society of the Pacific, p.727

\bibitem[Santos et al.(1995)]{setal95}
Santos Jr., J.F.C., Bica, E., Clari\'a, J.J., Piatti, A.E., Girardi, L.A., 
\& Dottori, H., 1995, MNRAS, 276, 1155

\bibitem[Santos \& Piatti(2004)]{sp04}
Santos Jr., J.F.C., \& Piatti, A.E. 2004, A\&A, 428, 79 (SP)

\bibitem[Schlegel et al.(1998)]{sfd98}
Schlegel, D., Finkbeiner, D., \& Davis, M. 1998, ApJ, 500, 525

\bibitem[Searle et al.(1980)]{swb80}
Searle, L., Wilkinson, A., \& Bagnuolo, W.G. 1980, ApJ, 239, 803

\bibitem[Silva \& Cornell(1992)]{sc92}
Silva, D.R., \& Cornell, M.E. 1992, ApJS, 81, 865

\bibitem[Stone \& Baldwin(1983)]{sb83}
Stone, R., \& Baldwin, J. 1983, MNRAS, 204, 347

\bibitem[Stothers \& Chin(1992)]{sch92}
Stothers, R. B., \& Chin, C.L. 1992, ApJ, 390, 136

\bibitem[Zaritsky et al.(1997)]{zht97}
Zaritsky, D., Harris, J., \& Thompson, I.B. 1997, AJ, 114, 1002

\end{thebibliography}
\end{document}